\documentclass[prb,twocolumn,amsmath,amssymb,showpacs] {revtex4}  
\usepackage{graphicx,color}
\newcommand{\al}{\alpha}

\newcommand{\de}{\delta}
\newcommand{\D}{\Delta}
\newcommand{\e}{\epsilon}
\newcommand{\g}{\gamma}
\newcommand{\s}{\sigma}
\newcommand{\del}{\nabla}
\newcommand{\p}{\partial}
\newcommand{\delt}{\partial_t}
\newcommand{\mbp}{\mathbf{p}}

\newcommand{\mbq}{\mathbf{q}}
\newcommand{\mbk}{\mathbf{k}}
\newcommand{\mbr}{\mathbf{r}}
\newcommand{\mbm}{\mathbf{m}}
\newcommand{\mbf}[1]{\mathbf{#1}}

\newcommand{\bfhat}[1]{\hat{\mathbf{{#1}}}}        

\newcommand{\mcal}[1]{\mathcal{#1}}
\newcommand{\tilA}{\tilde{\mbf {A}}}
\newcommand{\til}[1]{\tilde{#1}}

\newcommand{\bs}[1]{\boldsymbol{#1}}
\newcommand{\nn}{\nonumber\\}
\newcommand{\up}{\uparrow}
\newcommand{\down}{\downarrow}

\newcommand{\bea}{\begin{align}}
\newcommand{\eea}{\end{align}}
\newcommand{\ben}{\begin{equation}}
\newcommand{\een}{\end{equation}}
\newcommand{\bed}{\begin{displaymath} }
\newcommand{\eed}{\end{displaymath}}

\begin{document}
\title{Quantum kinetic equation in phase-space textured multiband systems}
\author{Clement H. Wong}
\author{Yaroslav Tserkovnyak}
\affiliation{Department of Physics and Astronomy, University of California, Los Angeles, California 90095, USA}
\date{\today}
\begin{abstract}
Starting from the density-matrix equation of motion, we derive a semiclassical kinetic equation for a general two-band electronic Hamiltonian, systematically including quantum-mechanical corrections up to second order in space-time gradients.  We find, in addition to band-projected corrections to the single-particle equation of motion due to phase-space Berry curvature, interband terms that we attribute to the nonorthorgonality of the projected Hilbert spaces. As examples, we apply our kinetic equation to electronic systems in the presence of spatially inhomogeneous and dynamical spin textures stemming from electromagnetic gauge potentials.   Specifically, we consider the electromagnetic response of massive two-dimensional Dirac fermions and three-dimensional Weyl fermions, and reproduce the anomalous currents known as the parity and the Adler-Bell-Jackiw anomaly in particle physics.
\end{abstract}



\maketitle
\section{introduction}
The semiclassical theory of electronic dynamics has been successful in explaining a wide range of transport phenomena in solid-state physics. It is well established that the semiclassical single-particle equations of motion (sEOM) acquire anomalous corrections that stem from Berry phases\cite{berryPRSLA84} accumulated in the adiabatic motion of a wave packet.\cite{karplusPR54,jungwirthPRL02,kohnPR57,adamsJPCS59,blountPR62,changPRL95,sundaramPRB99,haldanePRL04,culcerPRB05,shindouNPB05,xiaoPRL05,gosselinEPL06,duvalMPLB06,changJPCM08} Such Berry-phase effects have been successful in explaining the anomalous Hall effect,\cite{karplusPR54,jungwirthPRL02,haldanePRL04} corrections to semiclassical quantization,\cite{changPRL95,sundaramPRB99} intrinsic magnetic moments of electronic wave packets,\cite{sundaramPRB99,culcerPRB05,gosselinEPL06,xiaoPRL07,changJPCM08} and anomalous thermoelectric transport\cite{xiaoPRL06,zhangPRB09} in various electronic systems. The (noncanonical) Hamiltonian perspective and issues related to the Liouville's theorem and modified electron density of states have also been elucidated recently,\cite{xiaoPRL05,gosselinEPL06, duvalMPLB06} as well as Fermi-liquid generalizations.\cite{haldanePRL04,shindouPRL06}

Collective dynamics in the semiclassical approach is usually described with the Boltzmann equation, where electrons drift in phase space according to their single-particle equations of motion projected on respective bands.  While such an approach is intuitive and physically appealing, its validity is open to question.  The problem is that spatiotemporal inhomogeneities in the Hamiltonian such as electromagnetic potentials induce interband coherences that may not be completely captured by the sEOM.  Furthermore, it was noted in Ref.\,[\onlinecite{xiaoPRL05}] that the sEOM are noncanonical, appearing to violate Liouville's theorem.  As a remedy, the authors introduced a rescaling of the phase-space density of states.  This modification was shown to be consistent with a {formal requantization procedure} for band-projected Bloch electrons, which promotes the Poisson brackets of the sEOM to commutators, resulting in noncanonical commutator relations of position and momentum and the corresponding modified minimum quantum uncertainty in phase-space variables.\cite{xiaoPRL05}   However, we find no logical necessity to enforce Liouville's theorem for the band-projected distribution function,  and the formal requantization argument does not explain \emph{why} the density of states is modified by the band projection.  

These fundamental issues motivate us to consider the generalized Boltzmann equation for a multiband system from a systematic, ground-up approach.  In this paper, for a clean system with nondegenerate bands,  we derive a band-projected kinetic equation by performing a semiclassical expansion of the density-matrix equation of motion.  A similar Green's function approach was developed in Ref.\,[\onlinecite{shindouPRL06}].  However, there it was assumed from the outset that the distribution function was nonvanishing only in a single band.   A density matrix approach similar to the one to be presented in this paper was also used in Ref.\,[\onlinecite{culcerPRB06}] to analyze electronic transport on Bloch bands.  In this paper, we make no \textit{a priori} assumptions of decoupled bands and capture systematically all corrections to the Boltzmann equation up to second order in space-time gradients. Despite our attempt at decoupling by a systematic gradient expansion, we find remaining interband terms that warrant further investigation.

This paper is organized as follows.  In Sec.\,\ref{cov}, for a two-band Hamiltonian, we derive a covariant transport equation for the Wigner distribution function.  In Sec.\,\ref{adiab}, we decouple the transport equation to derive effective band-diagonal semiclassical kinetic equations (sKE).  In our sKE, in addition to the known Berry curvature corrections to the sEOM, we find terms corresponding to the aforementioned modified density of states, as well as interband terms not seen before.  We interpret these terms as representing quasiparticles motion in curved, nonorthogonal subspaces of the total Hilbert space.  In Sec.\,\ref{hydro}, we extract the hydrodynamic current from the continuity equation, discuss issues associated with momentum-space ultraviolet cutoffs and band-crossing points.  In Sec.\,\ref{EM}, we consider minimal coupling to electromagnetic gauge fields and express our transport equations in terms of the gauge-invariant kinetic momentum.  We then apply these equations to the 2D Dirac and 3D Weyl Hamiltonians in Sec.\,\ref{dirac2d} and Sec.\,\ref{weyl3d}.  
In the conclusion, Sec.\,\ref{con}, we qualitatively compare our approach with other methods for deriving band-projected effective Hamiltonians by canonical transformations on the Hilbert space.  Computational details are relegated to the appendices.   For the reader's convenience, we have provide references to the equations containing the main results of this paper in Table \ref{eqns}. 
\begin{table}[htdp]
\caption{Summary of the band-projected equations.}
\begin{center}
\begin{tabular}{l@{\quad}r@{\quad}l}
\hline
\hline
Equation & General & In electromagnetic fields\\
\hline
sEOMÊ & \eqref{eom} & \eqref{eom2r}, \eqref{eom2k}  \\
sKE & \eqref{pcont} & \eqref{kin1}\\
\hline\hline
``Anomalous" terms& Phase space& Hydrodynamic\\
\hline\hline
$F_{\mbp s}$ &\eqref{anom1}&\eqref{F0}\\
\hline
\end{tabular}
\end{center}
\label{eqns}
\end{table}%

\section{Semiclassical Kinetic equations}

\subsection {Covariant Transport Equation}
\label{cov}
Consider the many-particle density matrix,
 \ben
 \rho_{\alpha\beta}(\mathbf{r}_1,\mathbf{r}_2,t)\equiv\langle\psi_\beta^\dag( \mathbf{r}_2,t)\psi_\alpha(\mathbf{r}_1,t)\rangle\,,
 \een 
where $\psi_\alpha(\mathbf{r}_1,t)$ are second-quantized field operators, 
and $\alpha$ is a spin or band index.   We will consider only fermions in this paper, although our formalism is equally applicable to bosons.   For a quadratic, mean-field (or Fermi-liquid) Hamiltonian,
\ben
  H(t)=\int d\mathbf{r}_1d\mathbf{r}_2\,\psi^\dag_\alpha(\mathbf{r}_1,t)\mathcal{H}_ {\alpha\beta} (\mbf{r}_1,\mbf{r}_2,t) \psi_\beta( \mathbf{r}_2,t)\,,
 \label{H}
 \een
where $\hat{\mathcal{H}}$ is the first-quantized, quasiparticle Hamiltonian  expressed as a kernel (with the implied summation over repeated indices $\alpha,\beta$), the equation of motion for the density matrix closes: 
 \ben
  \partial_t\rho_ {\alpha\beta} + {i\over\hbar}\left(\mathcal{H}_ {\alpha\gamma}\otimes\rho_ {\gamma\beta}- \rho_ {\alpha\gamma}\otimes\mathcal{H}_ {\gamma\beta}\right) =0\,.
  \label{lou}
 \een
Here, $\otimes$ denotes a convolution integral in real space:  $[A\otimes B] (\mbr_1,\mbr_2)\equiv\int\,d\mbr'A(\mbr_1,\mbr')B(\mbr',\mbr_2) $.    We generally consider a smooth low-energy effective Hamiltonian
\ben
\mathcal{H}_{\alpha\beta}(\mbf{r}_1,\mbf{r}_2,t)=\epsilon_ {\alpha\beta} (-i\p_{\mbr_1},\mbf{r}_1,t)\de(\mbr_1-\mbr_2)\,.
\label{Hk}
\een
If necessary, the function $\epsilon_ {\alpha\beta} (-i\p_{\mbr_1},\mbf{r}_1,t)$ is properly symmetrized in its noncommuting arguments $\partial_{\mathbf{r}_1}$ and $\mathbf{r}_1$.  Eq.\,\eqref{Hk} may represent the continuum Hamiltonian for the slowly-varying envelope fields in the $\mbf{k}\cdot\mbf{p}$ expansion, or the continuum limit of a tight-binding Hamiltonian. 

For slow and long-wavelength spatiotemporal inhomogeneities, it is useful to define the distribution function $\hat{n}_\mbf{p}(\mbf{r},t)$ by the Wigner transform (WT) of the density matrix:\cite{reichlSP98}
\ben
\hat{n}_\mathbf{p}(\mathbf{r},t)\equiv\int\,d\mathbf{r}'e^{-{i\over\hbar}\mathbf{p}\cdot\mathbf{r}'}\hat{\rho}\left(\mathbf{r}+{\mathbf{r}'\over2}, \mathbf{r}-{\mathbf{r}'\over2},t\right).
\label{dist}
 \een 
 In the Wigner representation, the expectation value of an observable described by the kernel $\hat{A}(\mbr_1,\mbr_2)$ is given by
\ben
\langle\hat{A}\rangle(t)=\int{d\mbr}\sum_\mbp {\rm Tr}[\hat{n}_\mathbf{p}(\mathbf{r},t)\hat{A}_\mbp(\mbr,t)]\,,
\een
where $\sum_\mbp\equiv\int d^dp/(2\pi\hbar)^d$, $d$ being the number of spatial dimensions. Here and henceforth, we make a convention to denote the spin/band matrix structure by hats.  When $\hat{n}_\mathbf{p}(\mathbf{r},t)$ is a smooth function of $\mbr$ and $t$, one may construct a semiclassical kinetic equation for $\hat{n}_\mathbf{p}(\mathbf{r},t)$ by taking the WT of Eq.\,\eqref{lou} and performing a gradient expansion.  

The kinetic equation for $\hat{n}_\mathbf{p}(\mathbf{r},t)$ is governed by the quasiparticle energy matrix $\hat{\e}_\mbf{p}(\mbf{r},t)$, which is the WT of the quasiparticle Hamiltonian kernel (\ref{Hk}). 

We will go to a local spin frame that diagonalizes this semiclassical Hamiltonian, which for an n-band system, reads
\ben
\hat{U}_\mbp^\dagger\hat{\boldsymbol{\e}}_\mbp\hat{U}_\mbp=\rm{diag}(\e_1,\e_2,\ldots,\e_n)\,.
\een
where $\hat{U}_\mbp$ is a unitary transformation defined by this equation.
The basis defined by this rotation is the local energy eigenstates in the sense that the average total energy $\langle H \rangle$  may be expressed as a spatial integral (which defines the energy density)
\ben
 \langle H \rangle=\int{d\mbr} \sum_{\mbp s}\e_{\mbp s}n_{\mbp s}\,.
 \een  
$n_{\mbp s}$ here are the diagonal elements of the distribution function in the local spin frame.  
To derive the semiclassical kinetic equation, we take the WT of Eq.\,\eqref{lou} up to second order in the gradient expansion.  (See Appendix \ref{grad} for details.) 
The resulting kinetic equation in the local frame is
  \begin{align}
  D_t \hat {n}_\mbp + {i\over\hbar} [\hat{\epsilon}_\mbp, \hat {n}_\mbp] &- {1\over2}\{D_{i}\hat{\e}_\mbp,D^{i} \hat{n}_\mbp \}\nn
  &-{i\hbar\over8}[D_iD_j \hat{\e}_\mbp,D^iD^j \hat{n}_\mbp]=0\,,
  \label{kin}
  \end{align}
 where $\{,\}$ denotes anticommutator and $[,]$ commutator, $i,j$ label coordinates of phase-space vector $x^i\equiv(\mbr,\mbp)$, and summation over repeated indices is implied. The indices are raised by  $\p^i=J^{ij}\p_j$, where $\tensor{J}$ is the symplectic matrix acting on phase-space derivatives $\partial_i\equiv(\p_\mbf{r},\p_\mbf{p})$ as
 \ben
 \tensor{J} \left(\begin{array}{c} \p_\mbf{r} \\ \p_\mbf{p} \end{array}\right) = \left(\begin{array}{c} \p_\mbf{p} \\ -\p_\mbf{r} \end{array}\right)\,,~~~\tensor{J}=\left(\begin{array}{ll} 0 & 1 \\ -1 &0 \end{array}\right)\,,
 \een
the latter being written in terms of the $d\times d$ matrix blocks.\footnote{ Since $J^{ij}$ is antisymmetric, interchanging the positions of contracted raised and lower indices introduces an extra sign: 
 \[A_iB^i=A_iJ^{ij}B_j=-A_iJ^{ji}B_j=-A^iB_i\,.\]
 In particular,  $A_iA^i=0$.   Note that since $\tensor{J}^2=-{1}$, the inverse is $\tensor{J}^{-1}=-\tensor{J}$.}
In Eq.~(\ref{kin}), we introduced the covariant derivatives of the distribution function and energy defined by 
  \ben
  D_I\hat{M}\equiv\partial_I\hat{M}-i[\hat{A}_I,\hat{M}],
 \label{cd0}
 \een
 where the matrix-valued gauge fields entering covariant derivatives (\ref{cd}) are defined by $\hat{A}_I\equiv i\hat{U}_\mbp^\dag\p_I\hat{U}_\mbp$, and hereafter, the capital letters $I$, $J$,  and $K$ will be used to denote combined phase-space and time coordinates.  
 
 In the rest of this paper, for simplicity, we will consider a two-band system, then the energy matrix may be expanded as
 \ben
\hat{\epsilon}_\mathbf{p}(\mathbf{r},t)={1\over2}{\epsilon}_{\mathbf{p}} (\mathbf{r},t) +\boldsymbol{\D}_\mathbf{p}(\mbr,t)\cdot\hat{\boldsymbol{\tau}}\,.
 \label{qe}
 \een
Here, $\hat{\tau}_a=\hat{\s}_a/2$ are the spin-$1/2$ matrices satisfying $[\hat{\tau}_a,\hat{\tau}_b]=i\varepsilon^{abc}\hat{\tau}_c$ and $\{\hat{\tau}_a,\hat{\tau}_b\}=\delta_{ab}/2$,  $\hat{\s}_a$ being the Pauli matrices. $\varepsilon^{abc}$ is the antisymmetric Levi-Civita tensor, and we will use letters in the beginning of the alphabet for indices representing spin degrees of freedom. Below, we will be similarly decomposing the components of any $2\times2$ matrix into scalar and vector pieces as $\hat{M}=M/2+\mbf{M}\cdot\hat{\boldsymbol{\tau}}$, where $M\equiv{\rm Tr}[\hat{M}]$ and $\mbf{M}\equiv{\rm Tr}[\hat{M}\hat{\boldsymbol{\sigma}}]$. We parametrize the gap vector as $\boldsymbol{\D}_\mathbf{p}(\mbf{r},t)=\D_\mbp(\mbf{r},t)\mbf{m}_\mbp(\mbf{r},t)$, where $\D_\mbp=|\boldsymbol{\D}_\mathbf{p}|$ and $\mbf{m}_\mbp=(\sin\theta_ \mbp\cos\varphi_ \mbp,\sin\theta_ \mbp\sin\varphi_ \mbp,\cos\theta_ \mbp)$ is a unit-vector field represented by the spherical angles $\theta_\mbf{p}(\mbr,t)$ and $\varphi_\mbf{p}(\mbr,t)$.

The gap vector field $\boldsymbol{\D}_\mathbf{p}(\mbf{r},t)$ appears formally as a magnetic field coupled to spin in phase space, whose directional field $\mbf{m}_\mbf{p}(\mbf{r},t)$ we will call the \textit{spin texture}. Examples of Hamiltonian \eqref{qe} occur in ferromagnetic semiconductors, where the gap vector represents the exchange and spin-orbit fields, and in nonmagnetic semiconductors with spatially varying spin-orbit coupling. The internal degrees of freedom need not be the actual electron spin. For example, our kinetic equation can be applied to the pseudospin dynamics near the $K(K')$ points of graphene, which is described by the Dirac Hamiltonian.


We will be interested in quasiparticle transport on the semiclassical spin-orbit bands defined by the eigenvalues of Eq.\,\eqref{qe},
 \ben
 \e_{\mbp s}(\mbr,t) = {1\over2}\left[\e_{\mbp}(\mbr,t)+s{\D}_\mbp(\mbr,t)\right]\,.
 \label{disp}
 \een
 where $s=\pm1$. The local spin frame that diagonalizes Eq.\,\eqref{qe} is defined by an SU(2) rotation $\hat{U}_\mbp (\mbf{r},t)$ such that 
\ben
\hat{U}_\mbp^\dagger(\mbf{\D}_\mbp\cdot\hat{\boldsymbol{\tau}})\hat{U}_\mbp=\D_\mbf{p}\hat{\tau}_z\,.
\label{U}
\een
and the covariant derivatives in Eq. \eqref{cd0} may be written as
\ben
D_I\hat{M}={1\over2}\partial_IM+D_I\mbf{M}\cdot\hat{\boldsymbol{\tau}}\,,
 \label{cd}
 \een
 where $D_I\mbf{M} \equiv\partial_I\mbf{M}+\mbf{A}_I\times\mbf{M}$ and the vector-valued gauge fields are defined by $\hat{A}_I\equiv i\hat{U}_\mbp^\dag\p_I\hat{U}_\mbp\equiv\mbf{A}_I\cdot\hat{\boldsymbol{\tau}}=A_I^+\hat{\tau}_+ +A_I^-\hat{\tau}_-+{A}^z_I\hat{\tau}_z$, where $A_I^\pm=A^x_I\mp iA^y_I$ and $\hat{\tau}_\pm\equiv (\hat{\tau}_x\pm i\hat{\tau}_y)/2$.  In the Euler-angle  parametrization of our local spin frame, $\hat{U}(\varphi, \theta, \gamma)=e^{-i\varphi\hat{\tau}_z}e^{-i \theta \hat{\tau}_y}e^{-i \gamma\hat{\tau}_z} $,  the gauge fields are

\begin{align}
A_I^z&=\cos\theta \p_I \varphi+\p_I\gamma\,, \nn
A_I^{\pm}&=-e^{\pm i\g}(\sin\theta\partial_I\varphi\pm i\partial_I\theta)\,,
\label{gauge}
\end{align}
where $\gamma$ is an arbitrary rotation angle about $\mbf{m}_\mbf{p}(\mbf{r},t)$ and hence a local gauge parameter.    The form of $A_I^z$ reflects the north/south pole singularity in the spherical coordinate system, where $\varphi$ is not well defined. Near the poles, we may choose a gauge in which $\gamma=\mp\varphi$, which renders the gauge fields well behaved either at the north or south poles ($\theta=0$ or $\pi$), respectively, but not both.   It is thus necessary to use different gauges locally in regions where the texture passes through both north and south poles.  We emphasize that such singularities have a purely mathematical origin arising from our choice of coordinate system, and may occur where the texture is perfectly smooth.


The product of  the transverse components in Eq.\,\eqref{gauge} is a gauge-invariant second-rank tensor:
\begin{align}
A^+_I A^-_J=&\sin^2\theta\p_I\varphi\p_J\varphi+ \p_I\theta\p_J\theta \nn
&+i \sin\theta(\p_I\theta\p_J\varphi-\p_I\varphi\p_J\theta)\nn
\equiv&\mcal{G}_{IJ}+i \mcal{F}_{IJ}\,.
\label{tensors}
\end{align}
\begin{figure}[t]
\begin{center}
\includegraphics[width=.5\textwidth]{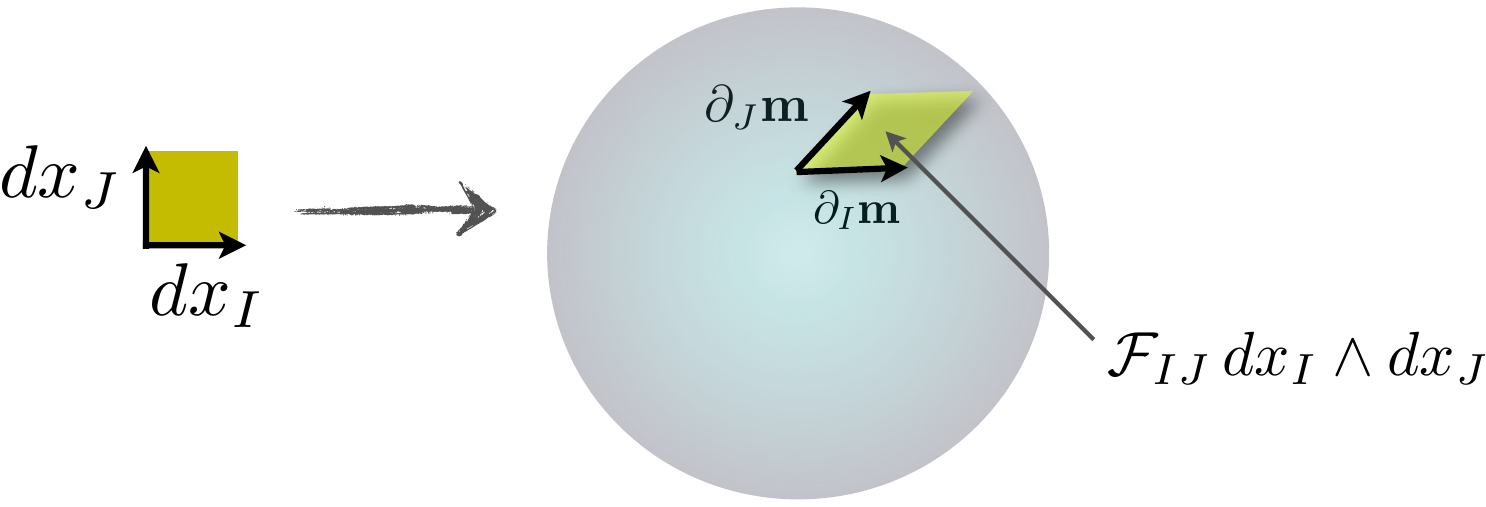}
\caption{In terms of the tangent vectors to the spin sphere,  $\p_I \mbf{m}=\p_I\theta \hat{\bs{\theta}}+ \p_I \varphi\sin\theta \hat{\bs{\varphi}}$, the tensors in Eq. \eqref{tensors} are $\mcal{G}_{IJ}=\p_i\mbf{m}\cdot\p_j\mbf{m}$,  $\mcal{F}_{IJ}=\mbf{m}\cdot(\p_i \mbf{m}\times\p_j \mbf{m})$, and geometrically represents the spin space metric and the area spanned by the tangent vectors, respectively. }
\label{tenfig}
\end{center}
\end{figure}

The real part $\mcal{G}_{IJ}$ is a kind of metric in spin space, which will not appear in the final results of this paper. 
\footnote{$\mcal{G}_{IJ}\equiv{\rm Re}\,A^+_I A^-_J$ is related to the field strength  by
\[\mcal{F}_ {IJ} =\varepsilon_ {IJ}\sqrt{(\det{\mcal{G}})_{IJ}}\,,\]
where we consider the metric tensor ${\mcal{G}_{IJ}}$ as a matrix in the $(x_I,x_J)$ plane and defined its determinant by $(\det{\mcal{G}})_{IJ}\equiv\mcal{G}_{II}\mcal{G}_{JJ}-\mcal{G}^2_{IJ}$.  This tensor appears in intermediate, gauge-dependent steps in our computations, but not in the final, gauge-invariant results.}
Fig. \ref{tenfig} illustrates the geometric meaning of these tensors.

\begin{figure}[t]
\begin{center}
\includegraphics[width=.5\textwidth]{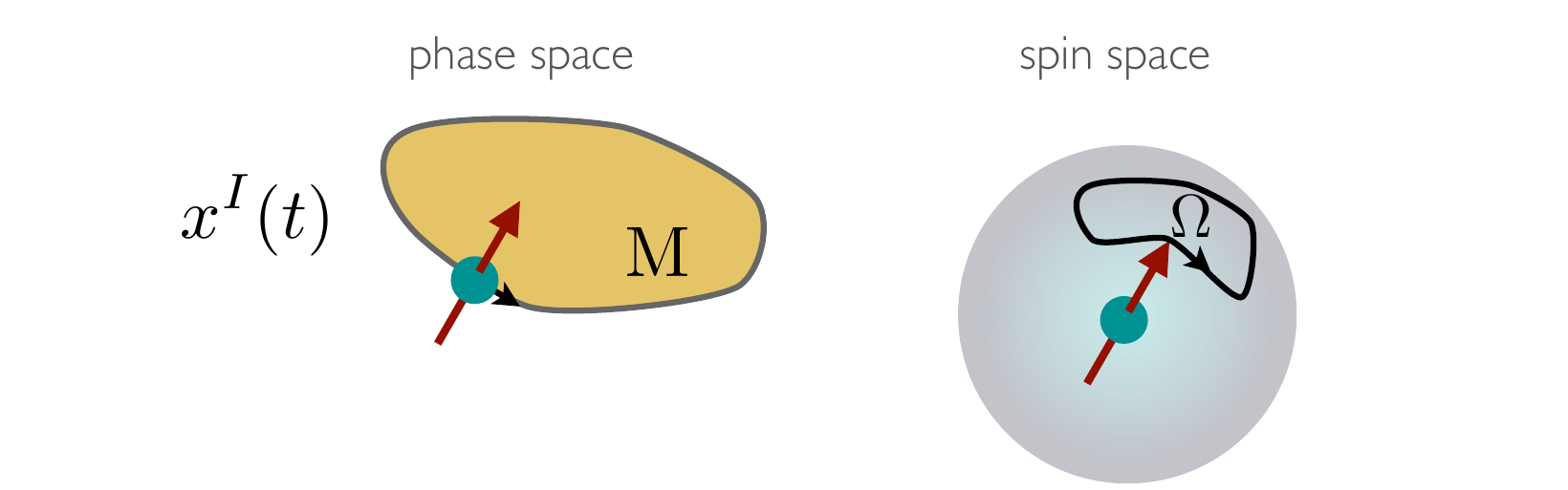}
\caption{Upon adiabatic transport around a loop $\p M$ which is the boundary of a hypersurface $M$ in the 7-dimensional phase space plus time, spin $s$ particles acquire a Berry phase $e^{iq_s\Phi}$, where $ q_s=-{s\over 2}$ and $\Phi=\oint_{\p M} dx_I\mcal{A}_I=\int_M \mcal{F}_{JK}dx^J\wedge dx^K={\Omega/2}$.}
\label{bphase}
\end{center}
\end{figure}

By gauge invariance, only the Berry curvature, i.e., the curl of the Berry gauge field $\mcal{A}_I\equiv -A^z_I$, appears in any physical quantities,
\ben
\mcal{F}_ {IJ}\equiv{\rm Im}\,A^+_I A^-_J= A^x_I A^y_J-A^y_I A^x_J=\p_I\mcal{A}_J-\p_J\mcal{A}_I\,.
\label{BC}
\een
In the rest of the paper, where necessary, we will denote the Berry gauge fields by $\mcal{A}^{\pm}_I=-(\cos\theta_ \mbp\mp1)\p_I \varphi_\mbp$ when well defined on the north/south pole, respectively. Geometrically, the Berry curvature gives the solid angle $\Omega$ spanned by the spin texture  $\mathbf{m}_\mbp(\mbr,t)$ per area in the $(x^I,x^J)$ plane. Nonvanishing Berry curvature means that particles acquire a phase-space Berry phase $(s/2)\oint dx_I\mcal{A}_I\equiv s\Omega/2$ over a closed trajectory, which modifies their transport.  (See Fig. \ref{bphase}.) All the phenomena we will investigate in this paper may be traced back to this phase.

Equation \eqref{kin} may be viewed as an expansion of Eq.\,\eqref{lou} in powers of $\hbar$.  However,  the separation into its classical $O(\hbar^0)$ and quantum $O(\hbar)$ part is not manifest because of its matrix structure, which represents the dynamics of quantum-mechanical internal degrees of freedom.  In the next section, we will derive a kinetic equation for the distribution function projected on each band defined by Eq.\,\eqref{disp}, which systematically captures all $O(\hbar)$ quantum corrections to the classical Boltzmann equation.

\subsection{Decoupling}
\label{adiab}
The diagonal part of kinetic equation \eqref{kin} reads 
  \ben
 \delt n_{\mbp s}+\partial^i\e_ {\mbp s}\partial_in_ {\mbp s} + F_ {\mbp s}(\p_i,\mbf{A}_i;{\e_s,n_s,\tilde{\mbf{n}}}) =0\,,
 \label{boltz}
 \een
where $F_ {\mbp s} $ are the $O(\hbar)$, ``anomalous" terms in the Boltzmann equation.  We denote transverse (i.e., $x,y$) vector components of the distribution {(in the local frame)} by $\tilde{\mbf{n}}$.   In the formulae throughout this paper, the partial derivatives $\p_i$ acts only on the symbol to its immediate right, so that in expressions of the form $\partial_iA\partial_jB$, $\p_i$ acts on $A$ only.
 At this point, the anomalous terms, which are shown explicitly in  Eqs.\,\eqref{F} and \eqref{Fz} of appendix \ref{decoupling}, may not appear manifestly gauge invariant (while they certainly should be).\cite{karplusPR54} Let us now decouple the longitudinal and transverse components in an adiabatic approximation, which will result in a closed, gauge-invariant equation for the diagonal distribution functions.   
The decoupling procedure may be organized in the following way.  Suppose that we can solve the transverse components in terms of the longitudinal  in a gradient expansion to the $(p-1) $th order in space-time derivatives: $\tilde{\mbf{n}}=\tilde{\mbf{n}}^{(0)}+\tilde{\mbf{n}}^{(1)}+\ldots+ \til{\mbf{n}}^{(p-1)}$.  By substituting  $\tilde{\mbf{n}}$  in $F_ {\mbp s} $, which is at least first order in gradients, and droppig any ($p+1$)th-order terms in Eq.\,\eqref{boltz}, we arrive at a $p$th-order equation for $n_{\mbp s}$. We will carry out this procedure to second order ($p=2$), consistent with our initial gradient expansion in Eq.\,\eqref{kin}.

From the off-diagonal part of Eq.\,\eqref{kin}, in regions where $\D\neq0$, one may readily find the transverse components to first-order space-time gradients [see appendix \ref{decoupling} for details]:
  \ben
\tilde{\mbf{n}}_\mbp={\hbar\over2}\left({n_{\mbp  z}\over\D}\partial_i\e_ \mbp-\p_in_ \mbp\right)\tilde{\mbf{A}}^i-\hbar{n_{\mbp  z}\over\D}\tilde{\mbf{A}}_t\,,
\label{tiln}
\een
where we have defined the phase-space particle-number and longitudinal spin densities,  $n= n_{\up}+n_{\down}$ and $n_z= n_{\up}-n_{\down}$, respectively.  The expression includes terms containing the expansion parameters of the adiabatic approximation: $(\hbar v_F/\D)\partial_\mbr$ and $(\hbar/\D)\p_t$, as well as a term $O(1/\D^0)$ that shows that interband coherences may be generated by gradients of the distribution, irrespective of the size of the gap and even in the absence of external fields.  We will see an example of this in the Dirac equation in Sec.\ref{dirac2d}. 

\begin{figure}[t]
\begin{center}
\includegraphics[width=.7\linewidth]{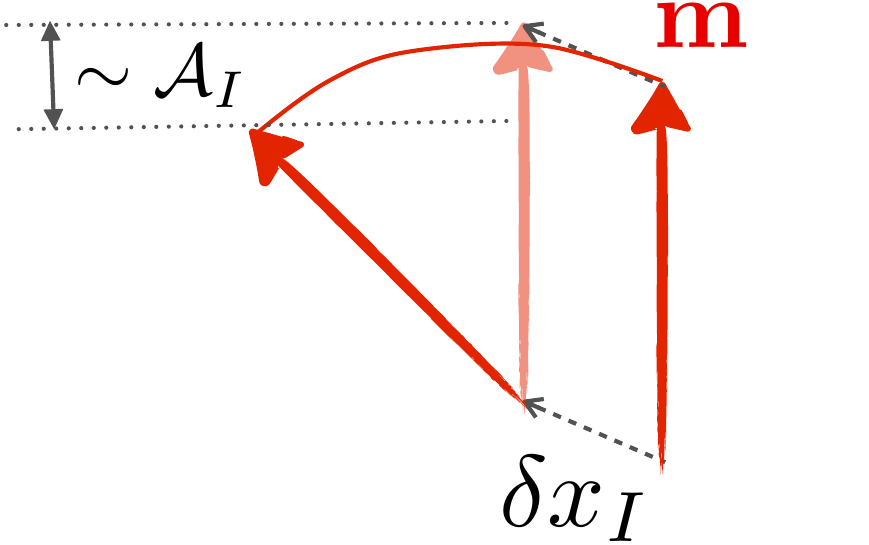}
\caption{This illustration shows the geometric meaning of the Berry gauge field.  It measures the misalignment between nearby spin ``up" states, where the local spin quantization axis is defined by $\mbm(x_I)$.}
\label{connect}
\end{center}
\end{figure}

Let us further consider why,  as shown in Eq.\,\eqref{tiln}, interband coherences (encoded by the transverse distribution function) are proportional to the transverse gauge fields.  Recall that the Berry gauge fields are the matrix elements ${A}_i^{ss'}=i\langle \mbf{m},s|\p_i | \mbf{m},s'\rangle$, where the local eigenspinors of our spin bands are given by  $|\mbf{m}_\mbf{p}(\mbr,t),s\rangle=\hat{U}_\mbp (\mbr,t)|\mbf{z},s\rangle$.   The transverse gauge connection determines the overlap between spin up and down states of nearby eigenspinors,  (while Berry gauge fields measures overlap between nearby up states, as illustrated in Fig. \ref{connect})
 \[\langle\mbf{m}(x,t),s| \mbf{m}(x+\de x,t+\de t),s'\rangle=\de^{ss'}-iA_I^{ss'}\de x^I\,.\]

 Consider the terms $O(n_z/\D)$ in Eq.\,\eqref{tiln}, in the presence of an electric potential in the energy.
The particle spins are locally misaligned with the texture $\mathbf{m}$ by an angle $\chi$ given by\footnote{The first term proportional to the electric field was first derived in the original paper of Karplus and Luttinger\cite{karplusPR54} on the anomalous Hall effect in ferromagnets.}
\[\tan\chi={|\til{\mbf{n}}|\over n_z}\to{{\hbar}\over\D}|E_l\til{\mbf{A}}_{p_l} {+}v_l\til{\mbf{A}}_{r_l}{+}\til{\mbf{A}}_t|\,,\] 
where the applied electric field is $\mbf{E}=-\p_{\mbf{r}}\e/2$, and we defined group velocity $\mbf{v}_\mbp=\p_\mbf{p}\e/2$. ($l$ runs over $d$ spatial dimensions.) A finite $\chi$ stems from the slight misalignment of electrons spin with the local quantization axis $\mbf{m}$ due to (i) drift in momentum space with velocity $\mbf{E}$, (ii) drift in real space with velocity $\mbf{v}_ \mbp $, and (iii)  dynamics of the texture.  In particular, a locally spin-up electron will have amplitude on the nearby down band proportional to $A_i^{+-}$. [See Eq.\,\eqref{tilA} for a microscopic expression for $A_i^{+-}$.]   For a near-equilibrium distribution function which depends only on energy, $n_ {\mbp s}(\e_{\mbp s})$, the  $O(1/\D^0)$ term reads $\p_\e n_ {\mbp s}\p_i\e\tilde{\mbf{A}}^i$.  Evidently, in this case, it also {stems from electron drift}, but comes from a different part of the phase space.  For a Fermi-Dirac distribution, this term originates from electrons on the Fermi surfaces, while the $O(n_z/\D)$ terms come from the momentum region between the Fermi surfaces.  

Substituting Eq.\,\eqref{tiln} in the anomalous terms in Eq.\,\eqref{boltz}, we find
\begin{align}
F_s=&\left[- {1\over4}\p^j(\D\mcal{F})+q_s(\mcal{F}_t{}^j-\mcal{F}^{ij}\p_i\e_s)\right]\p_jn_s\nn
&-\frac{1}{4}\left[\p^{i}\D\p_{i}n_{-s}+n_z(\p_t+\p^i\e_s\p_i)\right]\mcal{F}\,.
\label{anom1}
\end{align}
Here, the Berry curvature terms are expressed as 
\[\mcal{F}_{IJ}=\mbf{z}\cdot\mbf{A}_I\times\mbf{A}_J\,,\quad\mcal{F}\equiv\hbar\mcal{F}_{r_lp_l}\,,\]
recalling that index $l$ runs over $d$ spatial dimensions. We have defined the fictitious charge $q_s=-s\hbar/2$, $s=\pm1$ for the $\up,\down$ bands, respectively.  Equation \eqref{anom1} constitutes a central result of this paper.  It is clear that these are $O(\hbar)$ corrections to the Boltzmann equation \eqref{boltz}.  In the following sections, we will omit $\hbar$ for convenience (setting $\hbar=1$).   In the rest of this paper, we will explicate Eq.\,\eqref{anom1} and apply it to specific examples. 

The first line in Eq.\,\eqref{anom1} represents corrections to the single-particle equation of motion. Including only these terms, the longitudinal transport equation \eqref{boltz} would read
 \ben
 \p_tn_s-\left[(\p_i\bar{\e}_s+q_s\mcal{F}_{ti})J^{ij}+q_s\p_i\e_s\mcal{F}^{ij}\right]\p_j n_s+\ldots=0\,.
 \label{boltz1}
\een
We first note that there is a correction to the single-particle energy, $\bar{\e}_{\mbp s}=\e_ {\mbp s} +\de\e_ {\mbp s} $, where
\ben
\de\e_\mbp=-{\D_\mbp\mcal{F}\over4}\,,
\label{denergy}
\een
{which, in particular, is related to the magnetic moment of a semiclassical wave packet [see Eq.\,(\ref{Mk})].\cite{sundaramPRB99,changJPCM08}} The other Berry-curvature corrections introduce Hall-like terms to the single-particle equation of motion, which we \emph{define} so that the terms shown Eq.\,\eqref{boltz1} constitute the phase-space advective derivative of the distribution function, \[D^s_{t}n_s\equiv(\p_t+\dot{x}_s^j\p_j)n_s.\]   We thus identify phase-space velocites $\dot{x}_s^j=(\dot{\mbf{r}}_s,\dot{\mbf{p}}_s)$,\cite{sundaramPRB99,shindouNPB05, shindouPRL06}
 \begin{align}
 \dot{\mbr}_s&=\p_{\mbf{p}}\bar{\e}_s+q_s({\mcal{F}}_{t\mbp}+{\mcal{F}}_{\mbp p_l}\p_{r_l}{\e}_s -{\mcal{F}}_{\mbp r_l}\p_{p_l}{\e}_s )\,,\nn
 \dot{\mbp}_s&=-\p_{\mbf{r}}\bar{\e}_s-q_s ({\mcal{F}}_{t\mbr}+{\mcal{F}}_{\mbr p_l}\p_{r_l}{\e}_s -{\mcal{F}}_{\mbr r_l}\p_{p_l}{\e}_s)\,.
 \label{eom}
 \end{align}

As noted in Refs.\,\onlinecite{xiaoPRL05,gosselinEPL06,duvalMPLB06}, these equations of motion are noncanonical and thus appear to violate Louville's theorem.  Indeed, the phase-space velocity has a finite phase-space divergence,
\ben
\p_i\dot{x}_s^i
=q_s(\p_i\mcal{F}_t{}^i+\p_i\mcal{F}^{ik}\p_{k}\e_s)\,.
\label{div}
\een
From classical considerations of a two-component fluid, we would expect the projected kinetic equation to read: $\p_tn_{\mbp s}+\bs{\del}\cdot(n_ {\mbp s}\dot{\mbf{x}}_s)=\ldots$,
where we denote the phase-space gradient by $\bs{\del}={(\p_\mbr,\p_\mbp)}$, and $\ldots$ represents possible interband terms.  It may be seen, using the identity $\hbar\p^i\mcal{F}_{iJ}=\p_J\mcal{F}$ (see appendix \ref{bianchi}), that terms proportional to to Eq. \eqref{div} indeed appear in the second part of the second line in Eq.\,\eqref{anom1},  however, multiplied by the longitudinal spin density $n_z$ which couples the bands.\footnote{The appearance of interband couplings at the second-order gradient expansion is not inconsistent with the quantum adiabatic theorem.\cite{bransdenBOOK00} The latter allows transitions to other states to occur with a probability quadratic in the adiabatic parameter, which are space-time gradients in our problem. See appendix \ref{velocity}.}
To express our sKE as a kind of  phase-space ``continuity equation," we write $-n_z/4=(q_sn_s+q_{-s}n_{-s})/2= q_sn_s-q_{s}n/2$ in the expression for $F_s$, then our sKE reads
\ben
\p_tn_s+\p_i(n_s\dot{x}_s^i)=\left[\frac{1}{4}v^i_z\p_{i}n_{-s}+q_s{n\over2}(\p_t+\p^i\e_s\p_i)\right]\mcal{F}\,,
\label{pcont}
\een
where we have introduced notation for the relative band velocity: $v^i_{z}\equiv\p^i\D$.
Both sides of the equation contain terms that violate Louville's theorem.   Consider first the LHS.  Since the phase-space divergence is in fact the advective derivative of $q_s\mcal{F}$ (up to the order of our gradient expansion)
 \begin{align}
\p_i\dot{x}^i_s
&=q_s(\p_t+\p^i\e_s\p_i)\mcal{F}\nn
&=q_s(\p_t+\dot{x}^i_s\p_i)\mcal{F}+O(\p_\mu^3)
\equiv q_s{D}^s_t\mcal{F},
\label{DF}
\end{align}
we are lead to the following explanation for physical origin of the term $n_s\p_i\dot{x}^i_s$ on the LHS. 
In a semiclassical description, spinful particles occupy wave-packet states labeled by momentum and position, as well as a spin-coherent state labeled by spherical angles $\Omega$ on the unit sphere.  A particle is thus specified by a set of continuous coordinates, $(\mbr,\mbp,\Omega)$, which we will call the semiclassical \emph{state} space.
{
However, in our adiabatic approximation, we retained only part of the (matrix) distribution function, $n_{\mbp s}(\mbr,t)$,  which denotes particle occupation \emph{per} solid angle $\de\Omega$ subtended by the texture $\mbf{m}_\mbp(\mbr,t)$ in a volume $\prod_l\de r_l\de p_l$. Since this solid angle is determined by the phase-space Berry curvature, being for example in one spatial dimension  $\de\Omega (\mbp,\mbr,t) =\mcal{F}_{rp} \de r\de p$, the changes in the distribution function due to the modulation of this spin space volume along the phase-space particle trajectories appear in the transport equation \eqref{pcont} as an advective derivative of  $\mcal{F}\equiv{\hbar}\mcal{F}_{r_lp_l}$.     

In Ref.\,[\onlinecite{xiaoPRL05}], it was assumed that $n_s$ satisfies Eq.\,\eqref{pcont} with RHS equal to zero, in which case one could recover Louville's theorem for a rescaled distribution function  $f_s$ defined by  $n_s\equiv \mcal{D}_s f_s$,\cite{xiaoPRL05} where  $\mcal{D}_s$ satisfies $  \p_i\dot{x}^i_s =-(1/\mcal{D}_s)D_t \mcal{D}_s$, so that the LHS of Eq. \eqref{pcont} reads $\mcal{D}_sD^s_tf_s$. 
Since $\mcal{D}_s$ satisfies $\p_i\dot{x}^i_s=-D^s_t\ln\mcal{D}_s$, from Eq. \eqref{DF} we find
\ben
\mcal{D}_s=e^{-q_s\mcal{F}}\approx1-q_s\mcal{F}\,,
\een
Since we are neglecting terms that are cubic order in spatiotemporal derivatives at the level of the sKE, and, to the lowest order, $D_t\sim\p_\mu$, we should keep $\mathcal{D}_s$ only to linear order in gradients, as in the approximation above.  This expression agrees with the one quoted in Ref.\,[\onlinecite{xiaoRMP10}]. In Ref.\,[\onlinecite{xiaoPRL05}], $\mcal{D}_s$ was included as part of the phase-space measure representing a modified phase-space density of states for band-projected electrons.    Here, for our two band, continuum model, we have given an explanation for its origin.  It accounts for the (pseudo)spin degrees of freedom in an approximation in which  the quasiparticle spin dynamics is effectively constrained on a submanifold of the total state space.  This submanifold, which we will call the projected state space, is a hypersurface $\{\mbr, \mbp,\Omega (\mbr, \mbp,t)\}$ determined by the spin texture $\mbf{m}[\Omega(\mbf{r},\mbf{p},t)]$.
The projected state space has a local, dynamical volume proportional to $q_s\mcal{F}(\mbr,\mbp,t)$ and is thus \emph{curved}.   We note that this curvature vanishes,  $\mcal{F}=0$, unless   $\mbf{m}_\mbf{p}(\mbf{r},t)$ has both momentum and real-space derivatives. 

Now consider the terms in the RHS of Eq.\,\eqref{pcont}, which contain interband terms.  These terms are to be expected because the wave packets in the projected state space are not pointlike, but  occupy a finite minimum spin solid angle $\D\Omega$ corresponding to a finite minimum volume $\prod_{l}\D r_l \D p_l\sim\hbar^d$ in phase space required by the uncertainty principle.  However, due to the nonorthorgonality of the spin-coherent states, the spin up/down (along the texture) wave-packet states have nonzero overlap, resulting in interband couplings in the transport equation (See Fig. \ref{overlap}).  The overlap amplitudes are proportional to the area $\D\Omega\sim\mcal{F}$, consistent with the fact that the interband terms are proportional to $\mcal{F}$.
\begin{figure}[t]
\begin{center}
\includegraphics[width=.5\linewidth]{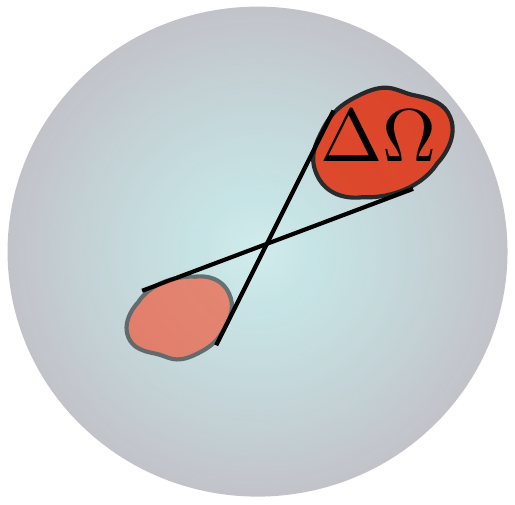}
\caption{Illustration of non-orthogonal areas on the spin sphere.}
\label{overlap}
\end{center}
\end{figure}


It is clear that these effects are generic to projected kinetic equation in multiband systems.  Since the decoupling procedure is technically much more difficult in this case, let us sketch out qualitatively how our two-band result might be generalized.  In a low energy, $\mbk\cdot\mbp$ expansion about a point (in the Brillouin zone) of high symmetry, electronic quasiparticle transforms under an irreducible spinor representation of the crystallographic point group,  which in some cases may be taken to be a higher-spin group, as in the case of the Luttinger Hamiltonian. The multicomponent electronic wave function may be viewed as an amplitude on a group manifold (in the representation space of the symmetry group).
Therefore, we may specify quasiparticles as in the two band case, except now $\Omega$ denotes coordinates on a higher dimensional group manifold, and Eq.\,\eqref{kin}, which governs coupled orbital and internal dynamics, may be viewed as quasiparticle motion in  $(\mbr,\mbp,\Omega)$.

For homogeneous spin-orbit couplings, the eigenfunctions $\hat{\psi}_s= \hat{u}_s(\mbp)e^{i\mbp\cdot\mbr}$ of the $\mbk\cdot\mbp$ Hamiltonian $\hat{\mcal{H}}(\mbp)$ defines a set of momentum-space spin textures.  When the spin-orbit field is inhomogeneous, we can still define a local, plane-wave-like (overcomplete) basis $\hat{\psi}_s= \hat{u}_s(\mbr,\mbp)e^{i\mbp\cdot\mbr}$,  where $\hat{u}_s(\mbr,\mbp)$ is an eigenvector of the Wigner-transformed Hamiltonian, and build wave packets out of these states.  Thus, the situation is quite similar to the two-band case, and we expect similar arguments pertaining to a curved state space and interband couplings due to the nonorthogonality of semiclassical coherent states to hold.    

 

\subsection{Hydrodynamics}
\label{hydro}
It is natural to expect a continuity equation for the particle density, $\rho=\sum_{\mbp}\text{Tr}[\hat{n}_{\mbp }]=\sum_{\mbp s}{n}_{\mbp s}$, which follows by taking $\sum_ {\mbp s} $ of our sKE's, resulting in an equation,
\[\partial_t\rho+\p_\mbr\cdot\mbf{j}+\ldots=0,\]
  from which we may extract the curl-free part of the particle current. 
 The {anomalous}, $O(\hbar)$ contribution to $\partial_t\rho$ is given by the terms in Eq.\,\eqref{anom1}, which, upon summation, may be written as [cf. Eq.\,\eqref{anom}]
\begin{align}
\sum_{\mbp s} F_ {\mbp s}
&=\sum_{\mbp s} \Big{\{}q_s\p_k\left[n_s(\mcal{F}_{it}J^{ik}-\p_i\e_{s}\mcal{F}^{ik} )\right]\nn
&\qquad\,\,\,\,\,\,-{1\over4}\p_k(\D n_s\p^k\mcal{F})\Big{\}}\nn
&=\sum_{s}\int {d^dp\over(2\pi)^d}  \left\{ \p_\mbr\cdot  \left[n_{\mbp s}\left(\de\dot{\mbr}_s -{\D \over4}\p_\mbp\mcal{F}\right)\right]\right.\nn
&\left.\qquad\,\,\,\,+\,\p_\mbp\cdot \left[ n_{\mbp s}\left(\de\dot{\mbp}_s+{\D\over4}\p_\mbr\mcal{F}\right)\right]\right\}\,,
\label{F0}
\end{align}
where we have denoted the Berry-curvature corrections proportional to $q_s$ in Eq.\,\eqref{eom} by $(\de\dot{\mbr}_s, \de\dot{\mbp}_s)$ (these are the corrections \emph{not} including the shifts in band energies $\p_i\de{\e}_s$). The momentum-space integration extends to a cutoff $p_\Lambda$ above which our low-energy effective theory is no longer valid.

From the first line in the second equality of Eq.\,\eqref{F0} and including the ``normal" part of the Boltzmann equation Eq. \eqref{boltz}, we identify the particle current 
\begin{align}
\mbf{j}=\sum_{s}\int& {d^dp\over(2\pi)^d} n_{\mbp s}[{{\p_\mbp}{\e}_s}-{\D\over4}\p_\mbp\mcal{F}\nn
&+ q_s({\mcal{F}}_{t\mbp}+{\mcal{F}}_{\mbp p_l}\p_{r_l}\e_s-{\mcal{F}}_{\mbp r_l}\p_{p_l}\e_s)]\,.
\label{current}
\end{align}
Note that {only part of the energy correction} in Eq.\,\eqref{denergy} and the corresponding contribution to group velocity enters into this current.  This resulted from cancelations which occurred in tracing the RHS of Eq. \eqref{pcont}.

The second line in Eq.\,\eqref{F0} appears as an anomalous particle source coming from the boundary in momentum space (invoking the divergence theorem).  Thus the $\ldots$ in our continuity equation reads,
\begin{align}
\sum_{s}\oint& {d^{d-1} S_\mbp\over(2\pi)^d}\mbf{l}\cdot\Big{\{}n_{\mbp s}\Big{[}-\p_{\mbf{r}}{\e}_ {\mbp s}+{\D\over4}\p_\mbr\mcal{F}\nn
&-q_s ({\mcal{F}}_{t\mbr}+{\mcal{F}}_{\mbr p_l}\p_{r_l}{\e}_s -{\mcal{F}}_{\mbr r_l}\p_{p_l}{\e}_s)\Big{]}\Big{\}}\,,
\label{bdcurrent}
\end {align}
where $\mbf{l}$ is the normal to the bounding surface at $p_\Lambda$.    This term represents particles flux coming from outside the momentum-space region where our semiclassical theory is valid.    In principle, if we know the microscopic theory beyond the cutoff and could solve for the distribution function there, we may use it as an input in our low-energy theory, providing a momentum-space boundary condition for our distribution function that is found below the cutoff.  In the following, for simplicity, we will take the cutoff to infinity.
Furthermore, it is well known that isolated degenerate points (band crossings) where $\D_\mbp=0$ are monopole sources of Berry curvature\cite{berryPRSLA84,haldanePRL04,murakamiPRL03} (see appendix \ref{bianchi}). These singularities occur because it is not possible to choose a spin frame [defined by the SU(2) rotation in Eq.\,\eqref{U}] that is continuous, near topological defects in the spin textures.\footnote{One should distinguish this from the case when $\D_\mbp =0$ but the texture is topologically trivial.} This occurs, for example, at the origin of the hedgehog texture in the Hamiltonian of Eq.\,\eqref{weyl}.  Strictly speaking, our expression for the transverse component in Eq.\,\eqref{tiln} is not valid at these points (since $\Delta=0$).\footnote{Furthemore, the Bianchi identity which was used in the derivation of Eq. \eqref{anom1} will acquire source terms at these points, cf. appendix \ref{bianchi}. } Generally, we will have to exclude these points by imposing a lower bound in our momentum integration defined by an infinitesimal surface bounding the singularity.   The boundary conditions for $n_{\mbp s}$ at these ``inner" boundaries may be found by solving the exact quantum-mechanical problem near the crossing point.


We note that since all terms $\propto \D^{-1}$ canceled out in the final transport equation, we may take the dc limit: $\omega\to0$, $\D\to0$ (in the prescribed order).  Thus in some instances,  one may regulate monopole singularities by introducing a gap in the Hamiltonian, compute the current, and then take the gap to zero.  However,  this procedure will in general introduce some ambiguity in the final answer.   For example, it may depend on the direction of the texture at the point where the gap is taken to zero.  A well-known example of this problem is the case of the massless 2D Dirac fermion,\cite{jackiwPRD84} where the vacuum current depends on the sign of the mass used to regulate the divergences for the massless fermion.  Below, we will work out the massive case in our semiclassical approach.

\subsection{Coupling to electromagnetic fields}
\label{EM}
Consider a situation where the real-space texture inhomogeneity stems \emph{only} from minimal coupling to vector and scalar gauge potentials, $\mbf{a}(\mbr,t)$ and $\phi(\mbr,t)$, in the first-quantized microscopic Hamiltonian, 
\ben
\hat{\mcal{H}}\left(-i\p_\mbr,\mbr,t\right)\to\mcal{H}\left(-i\p_\mbr-\mbf{a}(\mbr,t),\mbr,t\right)+\phi(\mbr,t)\,.
\label{em}
\een
The electron charge is absorbed here in the definition of the gauge potentials.

 From the Wigner transformation of this Hamiltonian, the semiclassical energy becomes  
 \[\e_\mbp\to\e_{\mbp-\mbf{a}(\mbr,t)}+\phi(\mbr,t)\equiv \e_\mbk(\mbr,t),\]
  where $\mbk (\mbr,t)\equiv\mbp-\mbf{a}(\mbr,t)$ is the kinetic momentum. The energy and the canonical momentum are not gauge invariant.   To make the transport equation manifestly gauge invariant, we must express the distribution function in terms of the kinetic momentum, $n_\mbp=n_{\mbk+\mbf{a}(\mbr,t)}(\mbr,t) \equiv n_\mbk (\mbr,t)$. Note that $\p_\mbk=\p_\mbp$, but the space-time derivatives $\p_\mu= (\p_\mbr,\p_t)$  in the transport equation are taken with fixed canonical momentum $\mbp$.  We have to take account of the implicit space-time dependence of kinetic momentum by
\ben
 \p_{\mu}|_\mbp=\p_\mu|_\mbk+\p_\mu\mbk\cdot\p_\mbk =\p_\mu-\p_\mu\mbf{a}\cdot\p_\mbk\,,
 \label{pmu}
\een
  where it is implicit now that $\p_\mu$ is taken with fixed $\mbk$. We will also impose a gauge invariant momentum-space cutoff $k_\Lambda$. 
   The transformation to kinetic momentum may be viewed as a phase-space coordinate transformation, under which the gradients transform as
 \begin{align}
\p_\mbr\e_{\mbp s}&=\p_\mbr\e_{\mbk s}-v_{\mbk s,l}\p_\mbr a_l\,,\nn
\p_\mbr n_\mbp&=\p_\mbr n_\mbk-\partial_{k_l}n_\mbk\p_\mbr a_l\,.
 \end{align}
Here, the group velocity is denoted by $\mbf{v}_{\mbk s}\equiv\p_\mbk{\e}_s$.  The transformation of the Berry-curvature corrections are given in appendix \ref{BCtrans}.  We find that $\mcal{F}=\mbf{b}\cdot{\boldsymbol{\mcal{B}}_\mbk}$,  so that the energy shift $\de \e_\mbk$ can be interpreted in terms of a magnetic moment $\boldsymbol{\mcal{M}}_\mbk$,
\ben
\de\e _\mbk
\equiv-\boldsymbol{\mcal{M}}_\mbk\cdot\mbf{b}
\label{Mk}
\een
where $\boldsymbol{\mcal{M}}_\mbk={\D}_\mbk\boldsymbol{\mcal{B}}_\mbk/4$. The factor of $1/4$ is a result of our conventions, where {we have factored out $1/2$ from the usual definition of the Berry gauge fields} in the literature. We show in appendix \ref{de} that this agrees with the magnetic moment derived more directly from wave-packet analysis.\cite{sundaramPRB99} The gauge-invariant transport equation now reads:
\begin{align}
&\left( \p_t+\dot{\mbf{r}}_s\cdot\p_\mbr+\dot{\mbf{k}}_s\cdot\p_\mbk \right) n_{\mbk s}\nn
&\quad-{n_z\over4}\left(\p_t+\mbf{v}_{\mbk s}\cdot\p_\mbr+\mbf{f}_{\mbk s}\cdot\p_\mbk\right)(\mbf{b}\cdot\boldsymbol{\mcal{B}}_\mbk)\nn
&\quad+ {\mbf{b}\cdot\boldsymbol{\mcal{B}}_\mbk\over4}\left(\mbf{v}_{\mbk z}\cdot\p_\mbr+\mbf{v}_{\mbk z}\times\mbf{b}\cdot\p_\mbk\right)n_{\mbk -s}=0\,,
\label{kin1}
\end{align}
where we have defined the relative velocity $\mbf{v}_{\mbk z}\equiv\mbf{v}_{\mbk \up}-\mbf{v}_{\mbk \down}$,  Lorentz force ${\mbf{f}}_s={\mbf{e}}+{\mbf{v}}_{\mbk s}\times\mbf{b}$, and ${\mbf{e}}\equiv-\p_\mbr{\phi}-\p_t\mbf{a}$ and $\mbf{b}\equiv\p_\mbr\times\mbf{a}$ are the ordinary electromagnetic fields.  In Eq. \eqref{kin1}, the phase-space velocities are {\cite{sundaramPRB99,shindouPRL06}}
\begin{align}
\dot{\mbf{r}}_s
&={\mbf{v}}_{\mbk s}+\p_\mbk\de\e_\mbk-q_s\mbf{f}_{\mbk s}\times\boldsymbol{\mcal{B}}_{\mbk}\nn
&=Z_{\mbk s}\mbf{v}_{\mbk s}+\p_\mbk\de\e_\mbk-q_s[\mbf{e}\times\boldsymbol{\mcal{B}}_\mbk+(\mbf{v}_{\mbk s}\cdot\boldsymbol{\mcal{B}}_{\mbk})\mbf{b}]\,,\label{eom2r}\\
 \dot{\mbf{k}}_s
 &={\mbf{f}}_s+\de{\mbf{f}}_s-q_s (\mbf{f}_{\mbk s}\times\boldsymbol{\mcal{B}}_{\mbk})\times\mbf{b}\nn
&=Z_ {\mbk s}{\mbf{f}}_s+\de{\mbf{f}}_s-q_s(\mbf{b}\cdot\mbf{e})\boldsymbol{\mcal{B}}_\mbk\,,
\label{eom2k}
\end{align}
where we define $Z_ {\mbk s} =1+q_s \mbf{b}\cdot\boldsymbol{\mcal{B}}_\mbk$ and 
\[\de{\mbf{f}}_s=-\p_\mbr\de\e_\mbk+\p_\mbk\de\e_\mbk\times\mbf{b}\,.\]
 We note that Eq.\,\eqref{eom2k} is not simply $\dot{\mathbf{p}}_s$ in Eq.\,\eqref{eom} transformed to kinetic momentum (which would not even be gauge invariant ), as it was necessary to transform the entire sKE.  In the second equalities for $ \dot{\mbf{r}}_s, \dot{\mbf{k}}_s$, we see that the group velocity and electromagnetic force are {rescaled and shifted.  The shift $\de\mbf{f}_{\mbk s}$ simply comes from the magnetic energy and the additional group velocity due to the magnetc moment in Eq.\,\eqref{Mk}}. The total $O(\hbar)$ correction to the group velocity, if we write {the first two terms} in the second equality for $ \dot{\mbf{r}}_s$ in Eq.\,\eqref{eom2r} as $\mbf{v}_{\mbk s}+\de\mbf{v}_ {\mbk s} $, is
\begin{align}
\de\mbf{v}_ {\mbk s}&=q_s(\mbf{b}\cdot\boldsymbol{\mcal{B}}_\mbk)\mbf{v}_ {\mbk s} +\p_\mbk\de\e_ {\mbk } \nn
&=-q_s (\mbf{b}\cdot\boldsymbol{\mcal{B}}_\mbk)(\mbf{v}_{\mbk}+s\mbf{v}_{\mbk z})-{\D_\mbk\over4}\p_\mbk(\mbf{b}\cdot\boldsymbol{\mcal{B}}_\mbk)\,.
\label{dv}
\end{align} 
 The third term in the first equality in expression for $\dot{\mbr}_s$ is the well-known  {anomalous Hall velocity}.\cite{adamsJPCS59} Furthermore, the last terms in the second equality of Eqs.\,\eqref{eom2r} and \eqref{eom2k} give an additional phase-space velocity in the direction of the phase-space ``magnetic fields."  

Up to the order of our gradient expansion, in which we need to keep terms in $\dot{\mbr}_s$(which multiplies $\p_\mbr n_s$) to linear order in EM fields, and in $\dot{\mbk}_s$ to quadratic order, and defining $\bar{\mbf{v}}_ {\mbk s}\equiv\p_\mbk\bar{\e}_ {\mbk s} $, $\bar{\mbf{e}}_{\mbk s} \equiv-\p_\mbr\bar{\e}_ {\mbk s}-\p_t\mbf{a}$ (recall that $\bar{\e}_{\mbk s}=\e_ {\mbk s} +\de\e_ {\mbk s} $), the equations of motion may be expressed as  
\begin{align}
\dot{\mbf{r}}_s&=\bar{\mbf{v}}_{\mbk s}-q_s{\dot{\mbk}}_s\times\boldsymbol{\mcal{B}}_\mbk\nn
 \dot{\mbf{k}}_s&=\bar{\mbf{e}}_ {\mbk s}+\dot{\mbr}_s\times\mbf{b}\,,
\label{eom1}
\end{align}
in agreement with the form of wave-packet equations of motion.\cite{sundaramPRB99}  We emphasize, however, that we are retaining terms beyond linear electromagnetic response\footnote{In the wave-packet formalism of Ref.\,[\onlinecite{sundaramPRB99}], the Hamiltonian is expanded to linear order in gradients, and therefore to linear order in electromagnetic fields.} which are not in the usual wave-packet equations of motion, for example in the $\mbf{b}\cdot\mbf{e}$ term for $\dot{\mbk}_s$ in Eq.\,\eqref{eom2k}, and includes forces such as $-\p_\mbr\de\e_\mbk$ coming from magnetic field gradients.

\section{examples}

\subsection{Massive 2D Dirac fermions}
\label{dirac2d}
 
As an application of our transport equation and a check on our formalism,  we consider the electromagnetic response of massive 2D Dirac fermions, which is relevant, for example, to the gapped surface state of  3D topological insulators.\cite{hasanRMP10,qiCM10}
 The Hamiltonian is given by 
 \ben
 \hat{\mcal{H}}_D= v\sum_{a=x,y} \hat{\s}_a[-i\p_\mbr-\mbf{a}(\mbr,t)]_a+mv^2\hat{\s}_z\,,
 \label{dirac}
\een
where $v$ is a constant with the units of velocity (which will be set to unity henceforth) and $\hat{\bs{\s}}$ is a vector of Pauli matrices.
 The particle/hole symmetric dispersions are conveniently expressed in terms of the relativistic energies $E_k \equiv\D_\mbk/2=\sqrt{\mbf{k}^2+m^2} $, so that $ \e_{\mbk s}=sE_k$ with corresponding group velocities  $\mbf{v}_{\mbk s}=s\mbf{k}/E_k$.  In $\mbk$ space, the texture is a vortex (meron) centered at the origin, $\varphi_\mbk=\arg(\mbk)$, with an out-of-plane component given by $\cos\theta_\mbk=m/E_k$. The vortex polarity is given by ${\rm sgn} (m)$. The spin texture has Berry curvature 
 \[\boldsymbol{\mcal{B}}_\mbf{k}=-\p_\mbk\cos\theta_\mbk\times\p_\mbk\varphi_\mbk=\frac{m}{E_k^{3}}\bfhat{z}\,.\]
The magnetic moment is thus $\mcal{M}_\mbk=m/2E_k^2$.   Both Berry curvature and magnetic moment are localized near the origin (see Fig.~\ref{bc}) and have a direction (normal to the $xy$ plane) depending on ${\rm sgn}(m)$.\cite{xiaoPRL07}
  \begin{figure}[h]
\begin{center}
\includegraphics[width=\linewidth]{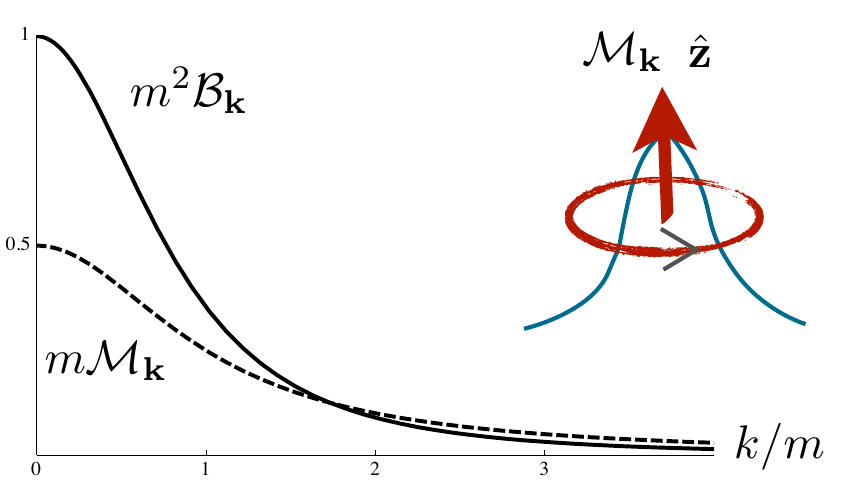}
\caption{In dimensionless variables, a plot of the Berry curvature, $m^2\mcal{B}_\mbk$ (solid curve), and magnetic moment, $m\mcal{M}_\mbk$ (dashed curve), as functions of $k/m$.   The magnetic moment stems from the orbital angular momentum of a finite size wave packet, as illustrated in the picture.}
\label{bc}
\end{center}
\end{figure}

Consider now applying a static, homogeneous magnetic field $\mbf{b}=b\bfhat{z}$.  From Eq.\,\eqref{dv}, the correction to the group velocity is $\de\mbf{v}_{\mbk s}=(bm/ 2E_k^4)\mbk$. The equations of motion Eq.\,\eqref{eom2r} and  Eq.\,\eqref{eom2k} then read
\begin{align}
\dot{\mbf{r}}_s
=\left(s+{bm\over 2E_k^3} \right){\mbk\over E_k}\,,\quad
 \dot{\mbf{k}}_s
 =b\dot{\mbf{r}}_s\times\bfhat{z}\,.
\label{cyc}
\end{align}
It is evident that the usual cyclotron motion holds, but the cyclotron orbits are not exactly traversed in the opposite sense for the two bands because of the $O(b)$ correction.  We also note that to reach this result, we had to keep the $O(b^ 2)$ term in the Lorentz force,  which arises due to the magnetic field dependence of the group velocity.  The second line in \eqref{kin1} vanishes on account of the spatially uniform static magnetic field and the fact that {$\mathbf{f}_{\mbk s}$ is transverse to $\mbk$}.  The third line reads
\ben
 {\mcal{B}\over2E_k^4}[b\mbk\cdot\p_\mbr+b^2\mbk\times\bfhat{z}\cdot\p_\mbk]n_{\mbk -s}\,.
 \label{interband}
 \een
For a spatially homogenous distribution (which is valid for an infinite plane),  the $O(b)$ term vanishes, and for a rotationally invariant distribution in $\mbk$, the $O(b^2)$ term vanishes.  Thus, in this case, the transport equation is simply solved by a constant distribution function.



To see some dynamics, consider adiabatically turning on the magnetic field, and let $b(t)$ be a slowly increasing function. We choose a rotationally-invariant vector potential that produces a uniform magnetic field along the $z$ axis, $\mbf{a}=b\bfhat{z}\times\mbr/2=b(-y,x)/2$, with the accompanying circulating electric field $\mbf{e}=-\p_t\mbf{a}=-\dot{b}\bfhat{z}\times\mbr/2$.
First, let us check that our equations agree with the St\v{r}eda formula,\cite{stredaJPC82} which is a general relationship between the Hall conductivity {in a gapped system} and the change in particle density $\rho$ when a magnetic field is turned on adiabatically.  For the rest of this section, let us restore $\hbar$, the speed of light $c$, and the electron charge by $\mbf{e}=e\mbf{E}$, $\mbf{b}=e\mbf{B}{/c}$. In 2D, the St\v{r}eda formula reads 
\[\s_{xy}=ec{\p \rho\over\p B_z}.\]

For a homogenous distribution function in linear response, the transport Eq.\,\eqref{kin1} reduces to
\begin{align}
\p_tn_{\mbk s}= {e\over2c}\sum_{s'} q_{s'} n^{(0)}_{\mbk s'}\p_t({\mbf{B}}\cdot\boldsymbol{\mcal{B}}_\mbk)\,,
\label{kin2}
\end{align}
 where $n^{(0)}_{\mbk z}$ is the unperturbed, longitudinal spin distribution that is zeroth order in $\mbf{B}$.  The resultant change in particle density $\de\rho$ for an adiabatic change $\de B_z$ is given by  
\begin{align}
ec{\de\rho\over\de B_z}
=\nu{e^2\over h}\,,
\label{drho}
\end{align}
where
\[\nu= \bfhat{z}\cdot\int d^2k\,{n}^{(0)}_{\mbk z}\mathcal{B}_\mbf{k}\,.\]
The extra particle density comes from particle fluxes at infinity due to the circulating electric field that produces anomalous Hall velocity $\dot{\mbf{r}}_{Hs}= -eq_s\mbf{E}\times\boldsymbol{\mcal{B}}_\mbk$. This results in particles on the upper (lower) band to enter (leave) the system. See Fig.\,\ref{streda}.
     \begin{figure}[t]
\begin{center}
\includegraphics[width=.4\textwidth]{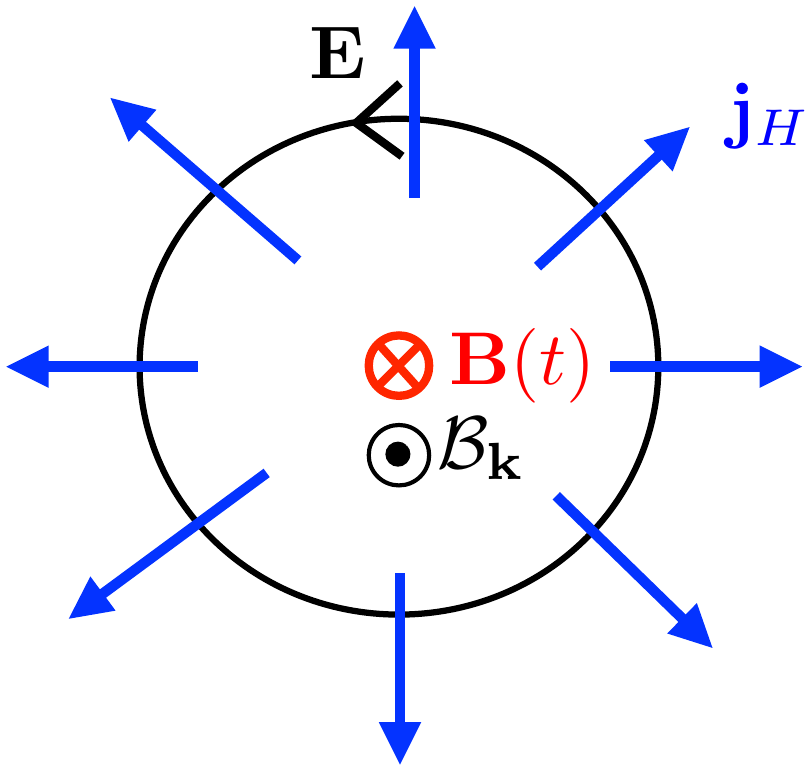}
\caption{With the chemical potential in the gap, the anomalous current (See Eq.\,\eqref{ac}) comes from particles in the lower band flowing out of the system due to the anomalous Hall velocity \eqref{hall}.}
\label{streda}
\end{center}
\end{figure}
One finds the corresponding Hall current
    \begin{align}
 \mbf{j}_H
 &=e\sum_{\mbf{k}s}\dot{\mbf{r}}_{H s}{n}^{(0)}_{\mbk s}
 =\nu{e^2\over h}\mbf{E}\times\mbf{z}\,,
 \label{hall}
\end{align}
and the conductivity $\s_{xy}=\nu{e^2/ h}$, which by inspection of Eq.\,\eqref{drho}, satisfies the St\v{r}eda formula.  We note that this is a general dynamical result, and, in particular,  does rely on assuming a Fermi-Dirac function for the unperturbed distribution.   We also emphasize here that according Eq. \eqref{kin2}, the St\v{r}eda formula for each band separately is {violated} and thus differs from the wave-packet theory.

If the unperturbed distribution function describes the ground state given by $n^{(0)}_{\mbk z}=\Theta(\e_F-\e_{\mbk\up})-\Theta(\e_F-\e_{\mbk\down})$,  then
 \ben
 \nu={\Phi_\up-\Phi_\down\over4\pi }\,,
 \label{nu}
 \een
  where $\Phi_s=\int {d^2k}\,  \Theta(\e_F-\e_{\mbk s})\,\mathcal{B}_\mbf{k}$  are the Berry curvature fluxes {over occupied regions} on each band.\cite{thoulessPRL82}  As another check, we show in appendix \ref{anomaly} that $\nu$ indeed gives the correct ground-state current of 2D Dirac massive fermions computed from field-theoretical methods.

Let us now consider our results in light of the well-known physics of the Dirac equation.  
In this case, the adiabatic approximation corresponds to the semiclassical limit, valid in the presence of weak external fields in which particle/antiparticle pair creation may be neglected.\cite{changJPCM08} Historically,  the single-particle interpretation of the Dirac equation showed some puzzling features.   The velocity operator given by $c\hat{\bs{\s}}$ in 2D has discrete eigenvalues $\pm c$ [restoring units here for the elementary electron, for which $v=c$ in Eq.\,\eqref{dirac}], which may seem to contradict the experimental fact that the electron's actual velocity is much less than the speed of light.  Futhermore, it has noncommuting components and, therefore, does not lend itself to a classical interpretation.   Dirac first resolved this apparent paradox by demonstrating that the electron trajectory exhibits a {trembling} motion called the ``Zitterbewegung:"\cite{diracBOOK47} The electron position moves with mean (group) velocity $\mbp c^2/E_\mbp$ and fast oscillations with frequency of the order of the mass gap $\D=2mc^2/h$ and amplitude of the order of the Compton wavelength $\lambda_c=h/mc$. However, when the electron's rest-mass frequency and Compton wavelength are beyond the finite time and spatial resolutions of real experiments, only the mean velocity and position are measured.  Our transport equation correctly captures the transport current due to the mean velocity. 


By analyzing the current of an electron wave packet, one can show that the group velocity $\mbp c^2/E_\mbp$ comes from the positive-energy band, while the Zitterbewegung comes from mixing with the negative energy band.\cite{itzyksonBOOK80}   Furthermore, it can be shown that the minimum size of a wave packet constructed from only positive-energy states is (in the nonrelativistic limit) of the order of $\lambda_c$.\cite{chuuSSC10} 
In other words, a semiclassical theory where the electron velocities are given by $\mbp c^2/E_\mbp$ requires a one-band description in which the size of electronic wave packets is much larger than $\lambda_c $, even in the absence of external fields.

This requirement shows up in our formalism as follows.   In the expression for the transverse distribution function $\til{\mathbf{n}}$ in Eq.\,\eqref{tiln}, the term $\sim\hbar\p_{r_l}n \til{\mbf{A}}_{p_l}$ must be small for our approximations to hold.  Consider the magnitude of transverse gauge field  in a relativistic expansion in the particle speed, $\beta_p\equiv|\mbf{v}_{\mbp s}|/c=pc/E_p=\sin\theta$,
\begin{align}
\hbar|\til{\mbf{A}}_{p_l} |&=\hbar\sqrt{\sin^2\theta(\p_\mbp\varphi)^2+ (\p_\mbp\theta)^2}\nn
  &=\hbar{\cos\theta}\sqrt {\tan^2\theta(\p_\mbp\varphi)^2+\cos^2\theta(\p_\mbp\tan\theta)^2}\nn
 &=\frac{\lambda_c}{2\pi}{\cos\theta}\sqrt{1+\cos^2\theta}=\frac{\lambda_c}{2\pi}\sqrt{(1-\beta_p^2)(2-\beta_p^2)}\nn
 &=\lambda_c\left[{\sqrt{2}\over 2\pi}+O(\beta_p^2)\right]\approx\lambda_c\,,
 \label{hA}
 \end{align}
where we used  $\tan\theta=p/mc$, $(\p_\mbp\varphi)^2=1/p^2$. 
 Thus we require 
 \[\left|{\til{n}_\mbp\over n}\right|\sim\left|{\hbar\p_{r_l}n \til{\mbf{A}}_{p_l}\over n}\right|\sim {\lambda_c\over l}\ll1,\]
  where $l$ is the lengthscale for spatial gradients of $n_\mbp$.  That is, the distribution function must be smooth on the scale of the Compton wavelength, consistent with the size requirement on nonrelativistic wave packets.  Note that in the ultrarelativistic limit, the right-hand side of Eq.\,\eqref{hA} approaches the de Broglie wavelength $h/p$, which then has to be much smaller than $l$. This is always required in the gradient expansion, even in the absence of interband terms.

The microscopic currents associated with the finite size of the electronic wave packet lead to an appearance of the magnetic moment [see Eq.\,\eqref{Mk}] $\boldsymbol{\mcal{M}}_\mbk=E_k\boldsymbol{\mcal{B}}_\mbk/2$.  In fact, the anomalous velocity may be understood from the relativistic physics of this magnetic moment, as follows.\cite{changJPCM08}   Under an applied $\mbf{E}$ field, in the frame of an electron moving with velocity $\mbf{v}_{\mbk s}=s\mbk c^2/E_k$, by the Lorentz transformation, there is a magnetic field given (for small velocities) by $\mbf{B}(\mbf{v}_{\mbk s})=-\mbf{v}_{\mbk s}\times\mbf{E}/c$. The magnetic energy in the electron rest frame then reads 
\begin{align}
\de\e_{\mbk s}=\frac{\hbar e}{c^2}\boldsymbol{\mcal{M}}_\mbk\cdot\mbf{v}_{\mbk s}\times\mbf{E}={\hbar e\over2}s\mbk\cdot\mbf{E}\times\boldsymbol{\mcal{B}}_\mbk\,,
\end{align} 
which in the lab frame may be interpreted as the energy of an effective electric dipole $\mbf{P}_{\mbk s}=(\hbar e/c^2)\mbf{v}_{\mbk s}\times \boldsymbol{\mcal{M}}_\mbk  $.  The contribution to the group velocity $\p_\mbk\de\e_{\mbk s}$ due to this extra energy is exactly the anomalous velocity.

Finally, we point out some surprising features in our band-diagonal transport equation that call for further investigation.  Equation \eqref{kin1} (where the second line is proportional to $n_z$) shows that the St\v{r}eda formula \emph{does not} hold separately on each band because half of the expected intraband flux goes into interband flux, which seems to suggest the following phenomenon.   Suppose the magnetic field smoothly falls off to zero in an outside region far away from the origin, where there are well-defined energy eigenstates.   Due to the anomalous Hall velocity, particles on the upper (lower) band flow in (out) of  the sample. Inside the region with magnetic field, particles are transferred between bands. If we started with the lower band occupied in the outside region, and slowly turn on the magnetic field, we might expect a pumping effect whereby some of the particles traveling into the origin would come out on the upper band (as defined by our WT basis). However, we note that the interband terms in the second and third lines of Eq.\,\eqref{kin1} that are induced by spatial inhomegeneities would also be present in this situation and need to be taken into account. The questions of whether the adiabatic interband fluxes in the presence of a finite gap can be physically manifested or if one could contrive a transformation that eliminates such interband terms in our transport equation \eqref{pcont} altogether remain open.

\subsection { Faraday's law in momentum space}
\label{far}
As another instructive example, consider a slowly-varying gap $m=m(t)$ of 2D Dirac fermions, which, for example, would occur due to a Zeeman splitting in the $z$ direction in topological insulators. For simplicity, we assume here the magnetic field is static. Then the transport equation is
\begin{align}
(\p_t+\dot{\mbf{r}}_s \cdot{\p_\mbr}+\dot{\mbf{k}}_s\cdot{\p_\mbk})n_{\mbk s}-{n_z\over4}(b\dot{\mcal{B}})=0\,,
\end{align}
where $\mcal{B}={m}/{E_k^{3}}$.  The time-dependent Berry flux  causes an anomalous velocity [cf. Eq.\,\eqref{eom}] $\de\dot{\mbf{r}}_s=q_s \mcal{F}_{\mbk t} \equiv q_s\bs{\mcal{E}}_{\mbk s} $,  where we have defined the momentum-space fictitious electric field which in this example is given by 
\[\bs{\mcal{E}}_{\mbk }=\p_t\cos\theta\partial_\mbk\varphi
={\dot{m}\over E_k^3}\bfhat{z}\times\mathbf{k}\,.\] 

This anomalous velocity is transverse to the cyclotron orbits and gives a radial particle flow in opposite directions on the two bands, similar to the previous example, except that it is peaked at $k=m/\sqrt{2}$.  This radial flow causes changes in particle density, determined by the time dependence of the Berry curvature: $\dot{\mcal{B}}=\dot{m}(k^2-2m^2)/{E_k^{5}}$. See Fig.\,\ref{vel}.  It may readily be verified the momentum-space, fictitious electromagnetic fields satisfies Faraday's law, $\p_t\mcal{B}_\mbf{k}+{\p_\mbk}\times\mcal{E}_\mbf{k}=0$ [See Eq.\,\eqref{df}].
 \begin{figure}[htbp]
\begin{center}
\includegraphics[width=.5\textwidth]{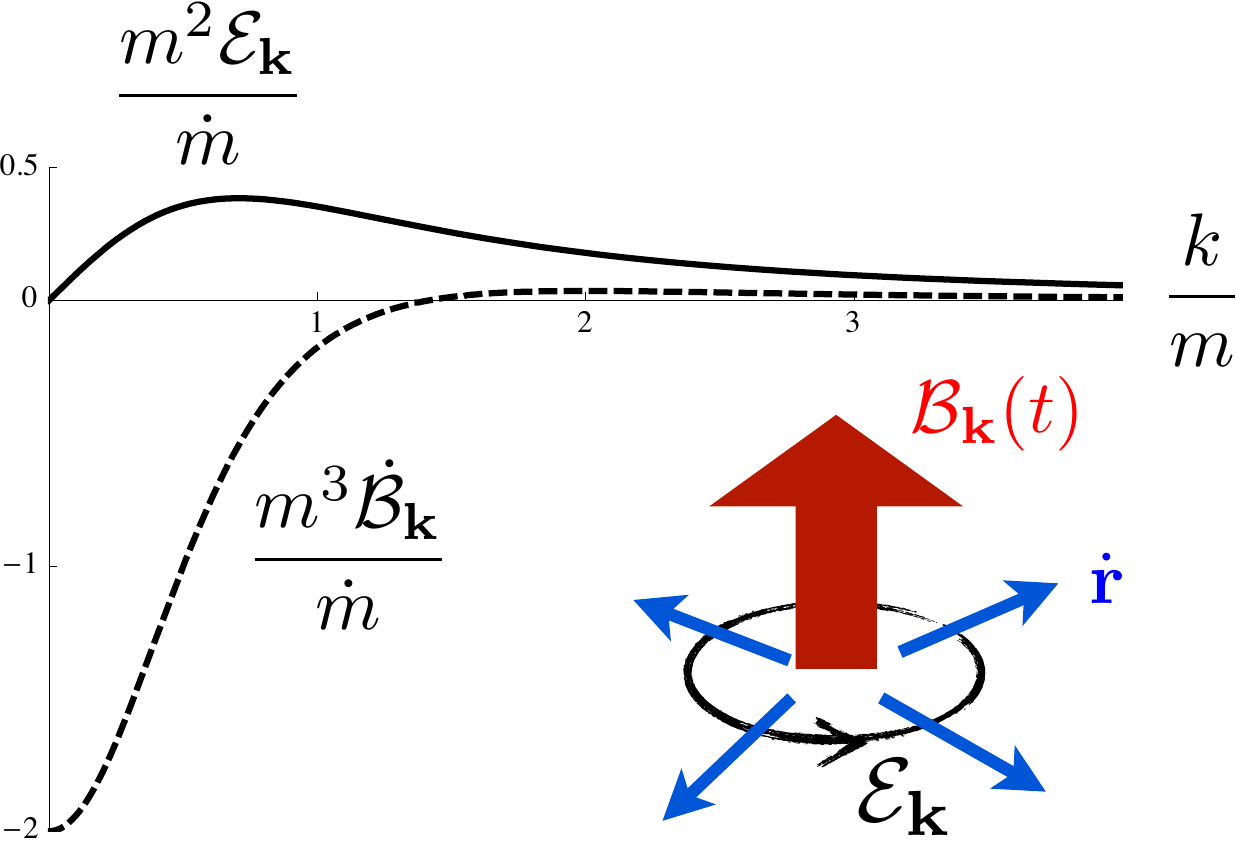}
\caption{A plot in dimensionless variables of the fictitious electric field, $(m^2/\dot{m})\mcal{E}_\mbk$ (solid curve), and Berry flux variation, $(m^3/\dot{m})\dot{\mcal{B}}_\mathbf{k}$ (dashed curve), versus $k/m$.}
\label{vel}
\end{center}
\end{figure}

\subsection{3D Weyl fermions}
\label{weyl3d}
As another illumimating example, we consider 3D Weyl fermions,  which represent 3D band crossing points.  In the presence of electromagnetic fields along the $z$ axis, the Hamiltonian for a right-handed (RH) Weyl fermion is
\ben
 \hat{\mcal{H}}_W= v\bs{\hat{\s}}\cdot\left[-i\p_\mbr-\mbf{a}(\mbr,t)\right]\,.
 \label{weyl}
\een
where $(a_x,a_y)$ is the vector potential of a magnetic field $b>0$ ($B_z<0$ for electrons) along  the $z$ axis, in the same gauge as in Sec.\,\ref{dirac2d}, and $a_z(t)$ gives the electric field $e_z=-\p_ta_z$.

The Hamiltonian in Eq.\,\eqref{weyl} has semiclassical dispersion $\e_\mbk=\pm|\mbk|$ and the Berry curvature of a monopole source at the origin:  $\boldsymbol{\mcal{B}}_\mbk=\mathbf{k}/k^3$, which causes a momentum-space particle flux represented by the $\mbf{b}\cdot\mbf{e}$ term for $\dot{\mbk}_s$ in Eq.\,\eqref{eom2k}.  As discussed in Sec.\,\ref{hydro}, to exclude the singularity at the origin, we must introduce an inner boundary in the momentum-space flux integral in Eq.\,\eqref{bdcurrent}. 
The particle fluxes through this boundary produce an anomalous source of particle density, (See  Fig. \ref{mono})
  \begin{align}
 \p_t\rho=\left[{e^2\over h^2} {1\over4\pi}\oint_{S^2} d\mbf{S}_\mbk\cdot\boldsymbol{\mcal{B}}_\mbk \, n_{\mbk z}\right]\mbf{B}\cdot\mbf{E}\,.
 \label{abj}
   \end{align}
   
 \begin{figure}[t]
\begin{center}
\includegraphics[width=.4\textwidth]{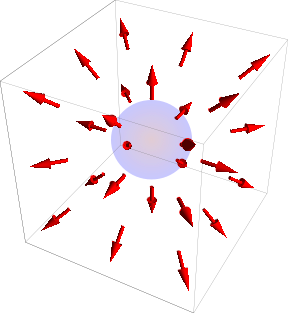}
\caption{Illustration of the momentum space texture of the Weyl equation, and the particle flux coming from the origin.}
\label{mono}
\end{center}
\end{figure}

Up to quadratic electromagnetic response, the integral in Eq.\,\eqref{abj} depends on the zeroth-order (in $\mbf{E}$ and $\mbf{B}$) distribution function $n^{(0)}_{\mbk z}$ on the boundary surface, which we take to be $S^2$. 
Assuming a ground-state distribution given by the Fermi level at zero energy, so that only the lower band is filled,  $n^{(0)}_{\mbk z}=-\Theta(\e_F-\e_{\mbk\down})$, the integral gives $4\pi$.   \footnote{Strictly speaking, the ground state is a Fermi sphere in the $\mbp$ space, and the corresponding distribution is slightly shifted by $\mbf{a}$ in the $\mbk$ space.} In units where $\hbar=1$, this results may be expressed as $\p_t\rho=-(e^2/4\pi^2)\,\mbf{B}\cdot\mbf{E}$, which is known in particle physics literature as the $(3+1)D$ Adler-Bell-Jackiw anomaly,\cite{nielsenPLB83} and, strictly speaking, cannot occur alone in nature because it violates charge conservation.  

 
Let us examine the source of this particle flux in the quantum mechanical solution of Eq.\,\eqref{weyl}.  In the presence of a magnetic field along $z$, the motion in the $x-y$ plane is quantized into Landau levels.  By translational invariance, the momentum along the magnetic field remains a good quantum number.   The spinor eigenfunctions are thus of the form $\Psi_{n}(\mbr)= e^{ip_zz}{\psi}_{p_z n}(x,y)$.  Since
 \[\hat{\mcal{H}}_W\Psi_{n}(\mbr)= e^{ip_zz}\hat{\mcal{H}}_D[m\to p_z]{\psi}_{p_z n}(x,y)\,,\]
  ${\psi}_{p_z n}(x,y)$ are the eigenfunctions of the Dirac Hamiltonian \eqref{dirac} with a mass $p_z$. The square of Hamiltonian \eqref{weyl} determines the eigenvalues (up to a sign), which are the relativistic Landau levels,\cite{haldanePRL88} 
 \[
E_{n\s}=\sigma v\sqrt{2|eB|n+p_z^2}\,,\,\,\,E_0=vp_z{\rm sgn}(eB)\,.
\]
Here $n=1,2, \ldots$ is a Landau-level index and $\s=\pm1$ labels the particle/hole branches.  The Landau levels spread out into bands in the $z$ direction, all of which are gapped, except for the ``zero mode" $E_0$, which has a linear, chiral dispersion along the direction of magnetic field. Starting with an equilibrium distribution with the Fermi level at zero energy, if we apply an electric field along the $z$ axis by adiabatically turning on $a_z$,  particles are transported above the Fermi level by spectral flow, locally populating states near the momentum-space origin.  These particles originate at $p_z=-\infty$ (outside the momentum-space cutoff).\footnote{Left-handed (LH) Weyl fermions [Eq.\,\eqref{weyl} with a minus sign] has a zero mode with the opposite chirality so that particles leave the system under spectral flow.  Therefore, Weyl fermions may occur in pairs with opposite chirality, so that particles are transferred between the crossing points in momentum space, conserving particle number.  This creates a nonequilibrium distribution in momentum space, which would relax back to equilibrium through scattering between crossing points.\cite{nielsenPLB83}}

The particle production rate may be computed from the rate at which the Fermi surface of the chiral branch changes, $\dot{p}_{Fz}=eE_z$, taking into account the zeroth Landau-level degeneracy.  The degeneracy of the Landau levels are given by $\Phi/\Phi_0$, where $\Phi=|B_zA|$ is the magnetic flux ($A$ is the transverse area) and, in units where $\hbar=1$ and $c=1$, the flux quantum is $\Phi_0=2\pi/e$.  Thus the density of states per area in the transverse direction is $eB/2\pi$, and per length along z is $1/2\pi$.  Therefore\cite{nielsenPLB83}}
\ben
\p_t{\rho}={eE_z\over2\pi}{e|B_z|\over2\pi}=-{e^2E_zB_z\over4\pi^2}\,.
\een
We may thus match this quantum-mechanical solution for $\p_t\rho$ with the semiclassical result given in Eq.\,\eqref{abj}.  We stress that our semiclassical analysis here involves response quadratic in electromagnetic fields and is, therefore, beyond the scope of the wave-packet analysis found in the literature.\cite{sundaramPRB99}

\section{Conclusion}
\label{con}

In summary, we have derived a band-diagonal, semiclassical kinetic equation (sKE) for electrons in a two-band,  (pseudo)spin-orbit coupled system, which takes the form of a Boltzmann equation [see Eqs.\,\eqref{boltz} and \eqref{anom1}] for a collisionless plasma.   In addition to the corrections to quasiparticle equations of motion,  we find terms proportional to the distribution function that we attribute to single-particle motion constrained on a curved state space.  We also find interband couplings that represent coherences due to the nonorthogonal nature of the projected Hilbert spaces.  As a check on our formalism, we find our kinetic equation reproduces the well-known electromagnetic response of 2D Dirac fermions and 3D Weyl fermions.   For Hamiltonians with less symmetry, more complicated inhomogeneous spin-orbit couplings, exact quantum-mechanical solutions are hard to find, while our sKE remains valid and a useful analytical tool.

At first sight, the interband terms in our sKE seem puzzling, given that there are well-defined procedures for decoupling multiband Hamiltonian in the presence of weak external fields.\cite{adamsJPCS59,blountPR62} However, these methods employ complicated quantum-mechanical transformations on the Hilbert space which generally do not constitute a simple change of basis.   Blount \cite{blountPR62} used a mixed representation where operators are specified by both Bloch (or canonical) momentum and position.  By successive transformations, that author derives an effective Hamiltonian for Bloch (Dirac) electrons in a magnetic field as a function of kinetic momentum $H(\mbk)$, which is a power series in $B$. The transformations are functions of $\mbk$ and become nonunitary in the presence of a magnetic field because of the commutator:  $[k_i,k_j]=i\e_{ijk}b^k$. As pointed out by the author, qualitatively, these transformations amount to a ``local" diagonalization of the Hamiltonian in phase space,  similar in spirit to our approach.  

For the {(3+1)D} Dirac equation, one may apply the Foldy-Wouthuysen (FW) transformation,\cite{foldyPR50}
 to decouple the positive/negative energy bands in the presence of weak external fields in a power series of inverse mass $1/m$, resulting in the nonrelativistic Pauli Hamiltonian.   The FW representation has the advantage that the operators retain their classical meaning.   In the absence of external fields, one can show that the position operator $\mbf{r}$ satisfies $\dot{\mbr}=\pm\mbp/E_\mbp$ (for the positive/negative energy bands),  and the orbital and spin angular-momentum operators are separately conserved.   However, these operators correspond to the original Dirac operators evaluated at the mean position of the electron, which, as discussed in Sec.\,\ref{dirac2d}, is spread out in a region of (at least) the size of the Compton wavelength $\lambda_c$.  In the presence of external fields, in principle, the meaning of the operators also changes and it may be seen explicitly that part of the FW transformation becomes nonunitary.

The key features of these methods are consistent with our formalism. Exact diagonalization of spin-orbit coupled Hamiltonian $\hat{\mcal{H}}(\mbp)$ is possible only in the absence of inhomogenieties. When the bands are gapped, one intuitively expects that in the presence of smooth space-time inhomogeneities, quasiparticle motion could be confined to separate subspaces, which may be called field-modified energy bands.\cite{adamsJPCS59}  One may indeed find representations in which the Hamiltonian is approximately block-diagonalized in the field-modified band space, but at the expense of altering the Hilbert space with nonunitary transformation, which may render the subspaces of the field-modified bands nonorthogonal and change the physical meaning of operators.  This caveat is consistent with our explanation of the interband terms in Sec.\,\ref{adiab}.  In cases where the Hilbert space may be specified by continuous coordinates, we find it useful to visualize the field-modified bands as submanifolds with possible nontrivial interband orthorgonality relations.

The Wigner distribution function gives us an exact phase-space representation of nonequilibrium quantum dynamics, where we can treat $\mbr,\mbp$ as real numbers from the get-go, allowing us to diagonalize the semiclassical Hamiltonian $\hat{\e}_\mbp(\mbr,t)$ in a straightforward manner.   Finding the exact correspondance between our \emph{matrix} transformation of the distribution function and the aforementioned transformation on the Hilbert space (where $\mbr,\mbp$ are operators) seems to be a nontrivial task, which we will relegate to future work.   Our approach based on the density-matrix formalism is quite general, allowing for the treatment of multicomponent fermions or bosons,  and in principle allows for a straightforward incorporation of electron interaction effects in the spirit of Fermi-liquid theory.  Lastly, our projection process described below Eq.\,\eqref{boltz} is valid up to higher orders, and, in particular, with some additional labor, one may  carry out the gradient expansion to 3rd order, which may warrant further study.\footnote{At this order, second-order derivatives of the diagonal distribution function will appear, and the transport equation cannot be interpreted in terms of continuous quasiparticle motion in phase space.}

\acknowledgements We are grateful to A. Kovalev and P. Krauss for stimulating discussions. This work was supported by the NSF under Grant No.~DMR-0840965 and DARPA.

\appendix

\section{Gradient expansion}
\label{grad}

The gradient  expansion expresses the WT of a convolution of two kernels in terms of gradients of the WT of the individual kernels through the following formula.\cite{rammerRMP86} Let $a(b)$ be the WT of $A(B)$, then WT of $[A\otimes B]$ is,
\begin{align}
{\rm WT}[A\otimes B]&={a}\left[ \exp{{i\hbar\over2}(\loarrow{\p}_\mbr\cdot\roarrow{\p}_\mbp-\loarrow{\p}_\mbp\cdot \roarrow{\p}_\mbr)}\right]b\nn&=ab+ {i \hbar\over2}\p_ia\p^ib- {\hbar^2\over8}\p_i\p_ja\p^i\p^jb+\ldots\,,
\label{WT}
\end{align}
where $\ldots$ indicate higher order terms.  For fermions, the gradient expansion is valid when the length scale $\xi$ of spatial inhomogeneities is much longer than the Fermi wavelength $\lambda_F\sim1/ q_F$,  where $q_F$ is the Fermi wave vector.  One may consider it as an expansion in $q/q_F$, and $q\sim1/\xi$ is the characteristic wave vector of the spatial inhomogeneities.
 
In fact, in the gradient expansion applied to  $A\otimes B-B\otimes A$  in Eq.\,\eqref{lou}, all even powers vanish if $A$ and $B$ commute because pairs of contracted indices are symmetric.  Therefore, in the scalar (single band) case, the leading quantum mechanical corrections are third order and $O(\hbar^2)$.\cite{wignerPR32}    However, when Eq.\,\eqref{lou} has nontrivial matrix structure, the second order expansion is required for capturing all semiclassical $O(\hbar)$  terms.   

\section{Covariance of the kinetic equation}

The transport equation \eqref{kin} is covariant in the sense that it remains the same form in an arbitrary spin frame.  Specifically,  under a local SU(2) transformation $\hat {n}_\mbp\to \hat {U}_ \mbp ^\dag \hat {n}_ \mbp \hat {U}_ \mbp $, Eq.\,\eqref{kin} represented symbolically as $d \hat {n}/dt$ transforms as $d \hat {n}/dt\to \hat {U}^\dag d \hat {n}/dt \hat {U}$.  The covariant derivatives transform similarly, with the appropriate change in the gauge potentials given by
  \begin{align}
  D_i\hat{n}&\to \hat {U}^\dag D_i\hat{n} \hat {U}\,,\nn
  \hat{A}_i &\to \hat {U}^\dag \hat{A}_i \hat {U} +i \hat {U}^\dag\p_i \hat{ U}\,.
   \end{align}
Expressing the matrix gauge potentials as $\hat {U}=\exp(i\eta^a \hat {\tau}^a)$, the infinitesimal transformation with $\eta^a\ll1$ is given by   
\begin{align}
\mbf{n}&\to\mbf{n}-\bs{\eta}\times\mbf{n},\nn
\mbf{A}_i&\to\mbf{A}_i-\bs{\eta}\times\mbf{A}_i+\p_i\bs{\eta}\,.
 \end{align}
Evidently, the spin components of the distribution function transforms in the adjoint representation of SU(2).

\section{Integrability condtion}

By their definition, the matrix Berry gauge potentials are pure gauge and so that their SU(2) field strength vanishes, which provides an integrability condition:  
\begin{align}
\p_i\mbf{A}_j-\p_j\mbf{A}_i&=-\mbf{A}_i\times\mbf{A}_j\,.
\label{int}
\end{align} 
This is a partial differential equation which relates the three component of the gauge field.  Mathematically, in a simply connected region where Eq. ~\eqref{int} holds, one can always find a frame in which the gauge potential vanishes.\cite{weinbergBook2}   Contracting indices in Eq. \eqref{int} gives an identity, 
\begin{align}
2\p^i\mbf{A}_i&=-\mbf{A}^i\times\mbf{A}_i\,,
\label{int1}
\end{align} 
which was used repeatedly to arrive at the final gauge-invariant expression in Eq.\,\eqref{anom}.
\section{Decoupling procedure}
\label{decoupling}

In this appendix, we show the computations that lead to the anomalous terms in Eq.\,\eqref{anom1} in projected kinetic equation.  It will be convenient to first express the longitudinal equation in terms of $n$ and $n_z$.  The trace and the $\hat{\tau}_z$ component of Eq.\,\eqref{kin} reads
 \begin{align}
 \p_t n&-{1\over2}(\partial_i\e\partial^in+\partial_i\D \partial^i{n}_z)+ F=0\,,\nn
 \p_tn_z&-{1\over2}[\partial_i\e\partial^i {n}_z+\partial_i\D\partial^i n ]+F_z=0\,,
\label{part}
 \end{align}
 where
 \begin{align}
 F\equiv&-{1\over2}\left[\partial_i\D(\tilde{\mbf{A}}^i \times \mbf{z} \cdot   \tilde{\mbf{n}})+{\D}(\mbf{z}\cdot\tilde{\mbf{A}}_i\times\p^i \tilde{\mbf{n}})\right.\nn
 &\left.\hspace{0.8cm}+{\D}\tilde{\mbf{A}}_i\cdot\tilde{\mbf{n}}{A}^{zi}\right]\,,
 \label{F}\\
  F_z\equiv&
 \mbf{z}\cdot\tilde{\mbf{A}}_t\times\tilde{\mbf{n}}-{1\over2}\p_i\e(\mbf{z} \cdot\mbf{A}^i \times\tilde{\mbf{n}})\nn
 &+{\hbar\over8}\mbf{z}\cdot\left(D^2_{ij}\boldsymbol{\e}\times D^{2ij}\mbf{n}\right)
 \label{Fz}
 \end{align}  
The vector part of the second covariant derivative,  defined by $ D_iD_j\hat{n}\equiv\partial_i\partial_jn /2+ D^2_{ij}\mbf{n}\cdot\hat{\boldsymbol{\tau}}$, reads
\begin{align}
D^2_{ij}\mbf{n}
 =&\p_i\p_j\mbf{n}+\p_i\mbf{A}_j\times\mbf{n}+\mbf{A}_j\times\p_i\mbf{n}\nn
 &+\mbf{A}_i\times\partial_j\mbf{n}+\mbf{A}_i \times(\mbf{A}_j\times\mbf{n})\,.
\label{cd1}
 \end{align}
Let us first simplify the last term of Eq.\,\eqref{Fz}.  The antisymmetric part of of the second covariant derivative Eq.\,\eqref{cd1} vanishes, as one would expect since the SU(2) field strength is zero.  Indeed, the antisymmetric part of the last term in \eqref{cd1} reads
 \begin{align*}
 (\mbf{A}_{[i}\times\mbf{A}_{j]} \times \mbf{n})^b
 &= {1\over2}[(\mbf{A}_i\times\mbf{A}_j)\times \mbf{n}]^b\,,
  \end{align*} 
   where we define anti-symmetrization by $A_{[i}B_{j]}\equiv (A_iB_j-B_iA_j)/2$.  Therefore, the commutator of covariant derivatives vanishes
 \ben
 [D_i,D_j]\mbf{n}=[\p_i\mbf{A}_j-\p_j\mbf{A}_i+\mbf{A}_i\times\mbf{A}_j]\times\mbf{n}\equiv\mbf{F}_{ij}\times\mbf{n}=0\,.
 \label{int3}
 \een
 where we've used the identity Eq.\,\eqref{int}. 
The remaining symmetric part of the covariant derivative Eq.\,\eqref{cd1} reads, 
 \begin{align}
 D_{(i}D_{j)}\mbf{n}&=\p_{ij}^2\mbf{n}+\p_{(i}\mbf{A}_{j)}\times\mbf{n}+2\mbf{A}_{(i}\times\p_{j)}\mbf{n}\nn
 &\quad+\mbf{A}_{(i}\cdot\mbf{n})\tilde{\mbf {A}}_{j)} -(\mbf{A}_i\cdot\mbf{A}_j)\mbf{n}\,,
 \label{cd2}
  \end{align}  
  where we defined the symmetrization symbol  $A_{(i}B_{j)}\equiv (A_iB_j+B_iA_j)/2$.   Therefore,  we will only need to compute that quantity
\ben
P^{(2)}_z
\equiv{1\over8}\mbf{z}\cdot\left(D^2_{(ij)}\bs{\e}\times D^{2(ij)}\mbf{n}\right),
\label{pz}
\een
of which we will only need the terms containing $n_z$, since the terms containing $\tilde{\mbf{n}}$ will be third order in gradients.  Taking $\mbf{n}\to n_z\bfhat{z}$,  noting that in the rotated frame we already have  $\bs{\e}=\D\mbf{z}$, and that the vector product \eqref{pz} above involves only the transverse components of the second covariant derivative \eqref{cd2}, we find

  \begin{align}
&{P}_z^{(2)}={1\over8}\mbf{z}\cdot\left\{\left[\D\p_{(i}\mbf{A}_{j)}+2\p_{(i}\D\mbf{A}_{j)}\right]\times\mbf{z}-\D\mcal{A}_{(i}\tilde{\mbf {A}}_{j)}\right\}\times\nn
&\qquad\left\{\left[n_z\p^{(i}\mbf{A}^{j)}+2\p^{(i} n_z\mbf{A}^{j)}\right] \times \mbf{z}-n_z \mcal{A}^{(i}\tilde{\mbf {A}}^{j)} \right\}\nn\,
&={1\over8}(\p_{i}\D n_z-\D\p_{i}n_z)\mbf{z}\cdot\tilA_{j}\times(\p^i \tilA^{j}+\p^j\tilA^{i}-\mcal{A}^j\mcal{G}_j{}^i) \nn
& \qquad+{1\over4}(\p_{i}\D\p^in_z\mcal{F}_j{}^j+\p_j\D\p^in_z\mcal{F}_i{}^j) \nn
&={1\over4}(\p_{i}\D n_z-\D\p_{i}n_z)\p^j\mcal{F}_j{}^i\nn
&\qquad-{1\over2}\p_{i}n_z\p^{i}\D\mcal{F}-{1\over4}\p_{i}n_z\p_{j}\D\mcal{F}^{ij} \,.
\label{P2}
\end{align}
In the third line, we use identity  \eqref{int1} and \eqref{bid1}.  This expression is gauge invariant, as it should be.
Next, we want to solve for $\til{\mbf{n}}$ to first order gradients.   Consider the transverse part of the transport Eq.\,\eqref{kin} in the first-order gradient expansion,
\ben
\tilde{D}_t \mbf{n}-{\D\over\hbar}\mbf{z}\times\tilde{\mbf{n}}-{1\over2} \left[\partial_i\e {\tilde{D}^i\mbf{n}}+\partial^in{\tilde{D}_i\bs{\e}}\right]=0\,,
\label{transverse}
\een
where the transverse covariant derivatives are defined by
\begin{align}
\tilde{D}_I\mbf{n}&= \partial_I\tilde{\mbf{n}}+{A}^z_I\mbf{z}\times\tilde{\mbf{n}}+n_z\tilde{\mbf{A}}_I\times\mbf{z}\,, \nn
{\tilde{D}_i\bs{\e}}&=\D {\tilde{\mbf{A}}_i\times\mbf{z}}\,.
  \end{align}

In regions where there is a finite gap, $\D_\mbp\neq0$, Eq.\,\eqref{transverse} shows that $\tilde{\mbf{n}}^{(0)}=0$, so that the lowest nonvanishing order is $\tilde{\mbf{n}}^{(1)}$.  Therefore, to first order in Eq.\,\eqref{transverse}  we can drop terms $\sim\p_I\tilde{\mbf{n}}, A^z_I\tilde{\mbf{n}}$, then
\begin{align}
{\D\over\hbar}&\mbf{z}\times\tilde{\mbf{n}}^{(1)}=n_z{\tilde{\mbf {A}}_t\times\mbf{z}}\nn
&-{1\over2}\left[\p_k\e (n_z\tilde{\mbf {A}}^k\times\mbf{z})+\partial^kn \D {\tilde{\mbf {A}}_k\times\mbf{z}}\right]\,.
\label{tiln0}
\end {align}
Solving for $\tilde{\mbf{n}}$ results in the solution Eq.\,\eqref{tiln}.

In approximating Eq.\,\eqref{transverse} with Eq. \eqref{tiln0},  we have neglected terms of $O(\p_t\til{\mbf{n}})$, $O(\til{\mbf{n}}\,\p_\mbf{p}\epsilon\cdot A^z_\mathbf{r},\til{\mbf{n}}\,\p_\mbf{r}\epsilon\cdot A^z_\mathbf{p})$, and $O(\p_\mbf{p}\epsilon\cdot\p_{\mbf{r}}\til{\mbf{n}},\p_\mbf{r}\epsilon\cdot\p_{\mbf{p}}\til{\mbf{n}}) $,   in comparison with the commutator term which is $O(\D\til{\mbf{n}})$.  Considering the near equilibrium case when $\tilde{\mbf{n}}$ is localized on the Fermi surface, our approximation implies the limit [see also Sec.~\ref{dirac2d} for additional constraint related to the second term in Eq.~\eqref{tiln}]
\ben
 \hbar\omega\ll\D\,,\quad \xi\gg{\hbar v_F\over\D}\,,\quad  \lambda_F|\p_\mbr\e|\ll\D\,,
 \een
 where $v_F$ is the Fermi velocity, $\omega$ ($\xi$) are the characteristic frequency (length scale) of the dynamics (inhomogeneities) of the system. The last condition on energy gradients is a requirement on the size of driving electromagnetic fields.  The limits above define the adiabatic approximation. Also, same conditions as above with $\Delta\to E_F$ ($\lambda_F\ll\xi$ etc.), for intraband adiabaticity.
 
Substituting the expression for  $\tilde{\mbf{n}}$ in \eqref{tiln0} into the anomalous terms of Eqs.\,\eqref{F} and  \eqref{Fz}, we find
\begin{align}
F
=&{1\over4} \mcal{F}^{ij}(\p_{i}\e\p_jn_z+\partial_i\D\partial_jn)-{1\over 2}\mcal{F}_t{}^i\p_i n_z\nn
&+{1\over4}(n_z\p_j\e-\D\p_j n)\p^i\mcal{F}_i{}^j-{1\over2} n_z\p^i\mcal{F}_{it}\nn
F_z
=&{1\over4}\mcal{F}^{ij}(\p_i\e\p_j n+\p_{i}\D\p_jn_z)-{1\over2}\mcal{F}_t{}^i\p_in \nn
&+ {1\over4}(n_z\p_j\D-\D\p_jn_z)\p^i\mcal{F}_i{}^j-{1\over2}\mcal{F}\p_{i}n_z\p^{i}\D\,.
\label{anom}
\end{align}
The anomalous terms in Eq.\,\eqref{anom1} are given by $F_ s =(F+sF_z)/2$. 

\section{Bianchi identity}
\label{bianchi}
The Bianchi identity for the Berry curvature reads
\ben
\p_I\mcal{F}_{JK}+\p_J\mcal{F}_{KI}+\p_K\mcal{F}_{IJ}=0,\quad I\neq J\neq K,
\label{bid}
\een
and follows directly from the definition of Berry curvature in terms of gauge fields if they are \emph {nonsingular}, since $\p_{[I}\mcal{F}_{JK]}=\p_{[I}\p_J\mcal{A}_{K]}\equiv0$.  Let $IJ=ij$ be phase-space indices and contracting identity \eqref{bid} by multiplying $\sum_{ij}J^{ij}$, then 
\ben
\p_i\mcal{F}^i{}_K+\p_K\mcal{F}=0\,.
\label{bid1}
\een
We have used \eqref{bid1} repeatedly, for example, we transformed Eq.\,\eqref{div} by,
 \begin{align}
\p_i\mcal{F}_t{}^i+\p_i\mcal{F}^{ik}\p_{k}\e_s=[\p_t+\p^i\e_s\p_i]\mcal{F}\,,
\label{dos}
   \end{align}
  We have used this identity in the second line of  \eqref{anom1}.

In the following,  it will be necessary to introduce differential forms in order to use Stokes theorem in a space with more than three dimensions.\cite{eguchiPRP80}   The Berry gauge field (connection) is a 1-form $\mathsf{A}= \mcal{A}_Idx^I$ and the Berry curvature is a two-form  $\mathsf{F}= {1\over2}\mcal{F}_{JK}dx^J\wedge dx^K$, where summation over repeated indices are implied, and the components of these forms are given in Eq. \eqref{tensors} and the text following Eq. \eqref{BC}. Denoting the exterior derivative by $d$, we may write  $\mathsf{F}={d} \mathsf{A}={d}(\cos\theta)\wedge{d}\varphi$, which is the area 2-form representing surface elements on the sphere.  The Bianchi identity \eqref{bid} states that  $d\mathsf{F}=0$ and represents a set of homogenous Maxwell equations for the phase-space, fictitious electromagnetic fields defined by 
\[\mcal{F}_{q_iq_j}\equiv\e_{ijl}\mcal{B}_{q^{l}}\,, \qquad   \mcal{F}_{\mbq t}\equiv\bs{\mcal{E}}_{\mbq}\,,\]
 where $\mbq\in\{\mbr,\mbp\}$ is a 3D vector in phase space which may have indices in both $\mbr$ and $\mbp$.  In differential forms notation, Eq. \eqref{bid} are the components of a 3-form, which may be expressed in terms of the fictitious electromagnetic fields, 
\begin{align}
d\mathsf{F}&={1\over2}\p_I\mcal{F}_{JK}\,dx^I\wedge dx^J\wedge dx^K\nn
&=\sum_{\mbq\in\{\mbr,\mbp\}}\left\{(\p_\mbq\cdot\mcal{B}_\mbq)\, dq^1\wedge dq^2\wedge dq^3\right.\nn
&\left.+{ \e_{lmn}\over2}(\p_t\mcal{B}_\mbf{q}+{\p_\mbq}\times\mcal{E}_\mbf{q})_{l}\, dt\wedge dq^m\wedge dq^n\right\}=0.
\label{df}
\end{align}
 Here $(lmn)$ runs through three dimensions. The first term represent the absence of monopoles and the second is the phase-space Faraday's law for the fictitious electromagnetic fields.

However, near crossing points, \eqref{bid} needs to be reconsidered,  since point degeneracies of the Hamiltonian are monopole sources of Berry flux\cite{berryPRSLA84} and Eq.\,\eqref{bid} assumes the absence of such singularities.  Consider first the Berry flux over a sphere containing the crossing point in phase space.   Modulo $4\pi$, it is an integer topological invariant known as  the Chern number: $\Phi=\int_{S^2} \mathsf{F}
=4\pi N$,  $ N\in\pi_2(S^2)=\mathbb{Z}$, and represents the winding number of the mapping between spherical surfaces from $(\mbr,\mbp,t)$ space to spin space defined by the texture, $\mbf{m}_\mbp(\mbr,t): S^2\to S^2$, analogous to the skyrmion number in ferromagnets.  Using Stokes theorem in the form $\int_M d \mathsf{F}=\int_{\p M} \mathsf{F}$, with $\p M=S^2$ being the surface of the sphere $M$,  we have 
\begin{align}
\Phi=\int_M d \mathsf{F}=4\pi N\,.
\label{flux2}
\end{align}
If we take the limit of $M$ being infinitesimally small and containing a topological defect of the texture at $\mbq_d$ in a 3D $\mbq$ space, only the first term of $d\mathsf{F}$ in Eq.\,\eqref{df} contributes to Eq. \eqref{flux2}, which implies that \cite{haldanePRL04}
  \begin{align}
\p_\mbq\cdot\mcal{B}_\mbq=\pm4\pi \de^3(\mbq-\mbq_d)\,.
   \end{align}
 Here we have used the fact that near a crossing point $\mbq_d$, one may make a linear expansion of the gap vector 
 \[\bs{\D}_\mbq =[(\mbq-\mbq_d)\cdot\p_\mbq]\bs{\D}_\mbq\equiv\sum_i(q^i-q_d^i)\mbf{e}_i,\]
 thus the texture has winding number $N=\pm1$, with the sign determined by $\det(e^a_i)$, where $e^a_i\equiv\p_i\D^a(\mbq_d)$ are a set of basis vectors defined by the gap gradients. The Weyl equation in Sec.~\ref{weyl3d} is an example of a monopole texture.
 
If, due to time-dependent parameters in the Hamiltonian, the monopole position changes in time, $\mbq_d=\mbq_d(t)$,  the current associated with the monopole motion will appear as a source term in Faraday's law.  To see this, consider the plane defined by $q^l=q^l_d(t_0 )$  in the 3D $\mbq$ space. Near $t_0$, we may approximate 
\[q^l-q_d^l(t)=(q^l-q^l_d(t_0))+\dot{q}_d^l(t_0)(t-t_0),\] 
Substituting this expression in the expansion above for the gap vector,  it is evident that there is a monopole texture located at  $(q_d^m,q_d^n,t_0)$ in $(q^m,q^n,t)$  space, and that the basis vector in the time direction is given by $\mbf{e}_t=\dot{q}_d^l\mbf{e}_l$.   The Berry flux over an infinitesimal surface containing this monopole in $(q^m,q^n,t)$ space, computed  using Eq. \eqref{flux2}, will have contributions only from the second term of $d\mathsf{F}$ in Eq. \eqref{df},  thus \cite{shindouPRL06}
\begin{align}
[\p_t\mcal{B}_\mbf{q}+&{\p_\mbq}\times\mcal{E}_\mbf{q}]_l\pm4\pi\e_{lmn}\nn
& \de(t-t_0) \de(q^m-q^m_d(t_0))\de(q^n-q^n_d(t_0))\,.
\label{faraday}
\end{align}
The sign here is determined by ${\rm sgn}[\dot{q}_d^l\det(e^a_i)]$, and LHS of Eq.\eqref{faraday} is evaluated at $q^l=q^l_d(t_0)$.  In terms of monopole motion, this nonzero winding number of the texture comes from the fact that, on the $q^l=q^l_d(t_0 )$ plane, the texture $\bs{\D}(q^m,q^n,q_d^l(t_0))$ is a vortex whose polarity, given by ${\rm sgn}(q_d^l(t_0)-q^l_d(t))$, switches sign during the course of time as the monopole passes the plane at $t=t_0$.  To express the RHS of Eq. \eqref{faraday}  in terms of the monopole current, we note that the linear expansion $q^l_d(t)$ near $t_0$ may be inverted to yield a mapping $t(q_d^l)$,  which we may use to change variables: $\de(t(q_d^l)-t_0)=|\dot{q}_d^l|\de(q_d^l(t)-q_d^l(t_0))$, and that on account of the delta function $\de(t-t_0)$, we may evaluated the LHS of Eq.\eqref{faraday} at $q^l=q^l_d(t)$.   Then,
\begin{align}
[\p_t\mcal{B}_\mbf{q}+&{\p_\mbq}\times\mcal{E}_\mbf{q}](t_0)={\rm sgn}[\det(e^a_i)]4\pi\dot{\mbq}_d\de^3(\mbq-\mbq^m_d(t_0))\,.
\label{faraday2}
\end{align}
Similarly, we note that the 2D Dirac equation (See Sec.\ref{far}) with a time-dependent mass $m(t)\propto t$ would give a source term in Faraday's law.


Lastly, the magnetic flux through an open 2D hypersurface $S$ may be converted to a contour integral of the vector potential around the boundary $\p M$, {provided that we use the appropriate gauge},
\begin{align}
\Phi
&=\int_{S}\mcal{F}=\oint_{\p S} dl^I \mcal{A}^{\pm}_I=-\int_0^1[\cos\theta(l)\mp1]\p_I \varphi (l) dl^I\,.
\end{align} 
The integral depends only on the path $[\theta(l), \varphi(l)]$ traced out by the texture in spin space space (which can be many closed loops) as the phase-space contour $\p M$ parametrized by $l\in[0,1]$ is traversed.   Except when the cone angle $\theta=\pi/2$, this flux is generally \emph{not} quantized.
  

\section{Berry curvature in general coordinates}
\label{BCtrans}

Under a general phase-space coordinate transformation $x^i\to\bar{x}^i(\mbf{x})$, the energy and the distribution function transforms as a scalar $\e(x)= \e(x(\bar{x}))\equiv\bar{\e}(\bar{x})$, while derivatives transforms as a covariant phase-space vector 
\ben
{\p\over\p x^j}\to\Lambda^i{}_j{\p\over\p\bar{x}^{i}},\quad\Lambda^i{}_j \equiv{\p\bar{x}^{i}\over\p x^{j}} \,.
\een
Since the Berry curvature is a product of derivatives, it transforms as a second rank covariant tensor  
\[\mcal{F}_{ij}=(\Lambda^T\bar{\mcal{F}}\Lambda)_{ij},\]
 where the bar denotes the Berry curvature with derivatives w.r.t. the coordinates $\bar{x}$.
In particular, when transforming to the kinetic momentum for electromagnetic perturbations, the spatial derivatives transforms as \eqref{pmu}, resulting in the following transformations of the Berry curvature 
\begin{align}
 \mcal{F}_{r^i t}&={\partial_i\mathbf{a}}\cdot{\partial_t\mathbf{a}}\times\boldsymbol{\mcal{B}}_\mbk\,,
  &\mathcal{F}_ {r^ir^j}&={\partial_i\mathbf{a}}\cdot{\partial_j\mathbf{a}}\times\boldsymbol{\mcal{B}}_\mbk\,,\nn
\mcal{F}_{p_it}&=-(\p_t\mbf{a}\times\boldsymbol{\mcal{B}}_\mbk)_i\,,&
 \mcal{F}_{p_i r_j}&=-(\p_j\mbf{a}\times\boldsymbol{\mcal{B}}_\mbk)_i\,,\nn
\mcal{F}_{p_ip_j}&=\mcal{F}_{k_ik_j}\equiv\e_{ijl}(\boldsymbol{\mcal{B}}_\mbk)_l\,, &
\mcal{F}_{r_i p_j}&=(\p_i\mbf{a}\times\boldsymbol{\mcal{B}}_\mbk)_j\,,
\label{bctransform}
 \end{align}
while $\mcal{F}$ transforms as
\ben
\mcal{F}=\sum_i \left({\p\mbf{a}\over\p r_i}\times\boldsymbol{\mcal{B}}_\mbk\right)_i=\sum_{ijl} \e^{ijl}{\p{a}^j\over\p r_i}(\boldsymbol{\mcal{B}}_\mbk)_l=\mbf{b}\cdot\boldsymbol{\mcal{B}}_\mbk\,.
\label{Ftransform}
\een
The simplest example of the transformations in Eq.\,\eqref{bctransform} is in the presence of an electric field in the vector potential gauge.  Then $\mcal{E}_\mbp=\mcal{F}_{\mbp t}=\mbf{e}\times\boldsymbol{\mcal{B}}_\mbk$ represents the anomalous Hall velocity.  This may be contrasted with the scalar potential gauge, in which the kinetic and canonical momentum are the same.

{In the presence of spatiotemporal inhomogeneities in the spin textures other than gauge potentials, the Berry curvatures will have space-time dependence not related to kinetic momentum, and there will be additional terms in the equations of motion Eq.\,\eqref{eom2r} and Eq.\,\eqref{eom2k}.  The additional terms come simply from the Berry curvatures in Eq.\,\eqref{eom} with space-time indices, with $\mbp\to\mbk$.
For example, if the texture is due to a ferromagnetic exchange field, the Berry curvatures $\mcal{F}_{\mbr\mbr}$ and $\mcal{F}_{\mbr t}$ are fictitious electromagnetic fields that are known to mediate spin-transfer torques in itinerant ferromagnets.\cite{wongPRB09}}

\section{Parity anomaly}
\label{anomaly}
For the example in Sec. \ref{dirac2d}, it is instructive to compute $\nu$ when the Fermi level lies in the gap, which gives the well known half-integer conductivity of 2D Dirac fermions.\cite{haldanePRL88}  Although the Berry flux in Eq.\,\eqref{nu} is concentrated near the origin, up to a sign they depend only on the texture far away from the origin.  To emphasize the topological nature of the number $\nu$, we apply Stokes theorem to convert integrals to Fermi surface Berry phases which determines the solid angle subtended by the texture, taking care to use the appropriate gauge, $\mcal{A}_\mbk=-(\cos\theta_ \mbk-\text{sgn}(m))\p_\mbk \varphi_\mbk$.   Then $\nu=-(1/4\pi)\oint\mcal{A}_\mbk\cdot d\mbk=-\text{sgn}(m)/2$,  in agreement with the formula well known in particle physics literature for anomalous charge current associated with the parity anomaly of the massive 2D Dirac equation,\cite{jackiwPRD84,semenoffPRL84}
  \ben
  \langle j^\mu\rangle=\text{sgn}(m){e^2\over8\pi}\e^{\mu\al\beta}F_{\al\beta}.
  \label{ac}
  \een  
In the formula above, $F_{\al\beta}$ is the electromagnetic field strength tensor, $F_{\mu\nu}=\p_\mu A_\nu-\p_\nu A_\mu$, $A_\mu=(\Phi,-\mbf{A})$, $F_{0i}=E_i$, $F_{ij}=-\e_{ijk}B^k$ , the expectation value is taken in the vacuum state, which in the single particle picture is the ``Dirac sea," with all negative energy levels occupied and thus corresponds to Fermi level in the gap.  One should  restore units in \eqref{ac} by setting $2\pi=h$ to compare with our result for $\nu$.  The dependence on $\text{sgn}(m)$ is a signature of the {infrared} singularity at $\mbk=0$ when $m=0$, as discussed in Sec.\,\ref{hydro}.  The 2D Dirac mass term explicitly breaks parity, defined by inversion of one spatial coordinate.  In field theory, even when $m=0$ in the Hamiltonian, gauge-invariant regularization of the {ultraviolet} divergence may violate parity (for example, in the Pauli-Villars regularization), leading to the parity-violating anomalous current.\cite{redlichPRL84}

{In field-theoretical methods, one calculates the anomalous current \eqref{ac} by the coupling of fermionic vacuum energy to an external vector potential, represented by a fermionic functional determinant.  While the result is precise, its physical origin is somewhat mysterious.    {The} semiclassical approach gives an intuitive but rigorous single-particle picture, in which the vacuum current simply represents the flow of quasiparticles.  

\section{Comparison with wave-packet theory}
\label{de}

According to wave-packet analysis,\cite{sundaramPRB99} the correction to the single particle energy due to inhomogeneous perturbations is given by,
\ben
\de\e_s={\rm Im}\,\langle\p_{r_l} u_s|\hat{H}_c(\mbr,t)-\e_s (\mbr,t)|\p_{p_l} u_{s}\rangle\,,
\label{d1}
\een
where $\hat{H}_c(\mbr,t)$ is the local Hamiltonian, with the position $\mbr$ evaluated at the center of the wave packet, and $\e_s (\mbr,t) $ is the band $s$ local eigenvalue.  We emphasize that in the wavepacket approach, $\mbr$ is a c-number parameter of a quantum Hamiltonian, whose complete set of eigenstates is known. The local Hamiltonian is the zeroth order term in an expansion of a Hamiltonian with smooth inhomogeneous perturbations.  The perturbations to first order in gradients has the form $\de H\sim(\hat{\mbr}-\mbr)\cdot\p_\mbr H$ and it's expectation value in a wave packet state gives the energy correction in first order perturbation theory.    From the semiclassical transport equation approach, the local Hamiltonian is equivilant to the semiclassical quasiparticle energy,
\[\hat{H}_c=\hat{\e}_\mbp(\mbr,t) \]
Next, we compare the formulae for our two band model with those  from wave-packet analysis.   First, we define the local eigenkets by a rotation of the lab frame spin eigenkets pointing in the $\pm\bfhat{z}$ direction: \[|u_{\mbp s} (\mbr,t)\rangle\equiv U_\mbp(\mbr,t) |\bfhat{z};s\rangle\,.\]  They satisfy the local eigenvalue equation:
 \ben
 [\hat{H}_c (\mbr,t)-\e_s (\mbr,t)] |u_{\mbp s}(\mbr,t)\rangle=0\,,
 \label{D1}
 \een 
where $\e_s$ is given by \eqref{disp}.   For brevity, we will denote ``lab" frame spinor basis by $|s\rangle$, $s=\pm1$.  The matrix elements of the gauge fields are given by 
\ben
[\hat{\mbf{A}}_i\cdot\hat{\bs{\tau}}]^{ss'}\equiv\hat{A}_i^{ss'}=i\langle s| \hat {U}^\dag\p_i \hat {U}|s'\rangle=i\langle u_s|\p_iu_{s'}\rangle\,,
\label{b1}
\een
where we've omitted phase-space subscripts for brevity. From \eqref{d1}, we find 
\begin{align}
\de\e_s&= {1\over2i}\langle s|\p_i U^\dag[{\D}\mbf{m}_\mbp\cdot\hat{\bs{\tau}}-s\D/2]\p^i U|s\rangle  \nn
&= {1\over2i}\langle s|\p_i U^\dag U[{\D}\hat{\tau}_z-s\D/2]U^\dag \p^i U|s\rangle\nn
&= {1\over2i}\left[\hat{A}_i {\D\over2}
\left(
\begin{array}{cc}
1-s  &   0\\
0  & -1-s    
\end{array}
\right)
 \hat{A}^i\right]_{ss}\nn
 &={\D\over2i} \sum_{s'\neq s}{s'\over2} \hat{A}^{ss'}_i\hat{A}^{{s's}i}\nn
 &={\D\mcal{F}\over4}
\end{align}
where $\mcal{F}\equiv\mcal{F}_i{}^i/2$.  Note that the energy shift is independent of $s$. 

\section{Relation of gauge fields to velocity operator}
\label{velocity}

We note that the off-diagonal gauge field is related to the phase-space velocity operator, $\hat{v}_i=\p_i\hat{H}$.  Differentiating the eigenvalue equation \eqref{D1} and projecting on $|u_{s'}\rangle$ for $s\neq s'$, we find an equation for the off-diagonal gauge fields [cf. \eqref{b1}]: 
\ben
{{A}}_i^{ss'}=-i\frac{\langle u_s|\p_i\hat{H}|u_{s'}\rangle}{\e_s-\e_{s'}}\,.
\label{tilA}
\een
Consider going into the frame of the electron moving in phase space, then the phase-space dependence of the texture appears to the electron as dynamics of a local phase-space magnetic field.   Therefore, to make heuristic comparison with the quantum adiabatic theorem,\cite{bransdenBOOK00} we take $\p_i\to\p_t$ in Eq.\,\eqref{tilA}. Then the theorem shows that the amplitude for transitions to other states is given by ${{A}}^{ss'}_i/\D$ in agreement with Eq.\,\eqref{tiln}.


\begin{thebibliography}{38}
\expandafter\ifx\csname natexlab\endcsname\relax\def\natexlab#1{#1}\fi
\expandafter\ifx\csname bibnamefont\endcsname\relax
  \def\bibnamefont#1{#1}\fi
\expandafter\ifx\csname bibfnamefont\endcsname\relax
  \def\bibfnamefont#1{#1}\fi
\expandafter\ifx\csname citenamefont\endcsname\relax
  \def\citenamefont#1{#1}\fi
\expandafter\ifx\csname url\endcsname\relax
  \def\url#1{\texttt{#1}}\fi
\expandafter\ifx\csname urlprefix\endcsname\relax\def\urlprefix{URL }\fi
\providecommand{\bibinfo}[2]{#2}
\providecommand{\eprint}[2][]{\url{#2}}

\bibitem[{\citenamefont{Berry}(1984)}]{berryPRSLA84}
\bibinfo{author}{\bibfnamefont{M.~V.} \bibnamefont{Berry}},
  \bibinfo{journal}{Proc. R. Soc. London A} \textbf{\bibinfo{volume}{392}},
  \bibinfo{pages}{45} (\bibinfo{year}{1984}).

\bibitem[{\citenamefont{Karplus and Luttinger}(1954)}]{karplusPR54}
\bibinfo{author}{\bibfnamefont{R.}~\bibnamefont{Karplus}} \bibnamefont{and}
  \bibinfo{author}{\bibfnamefont{J.~M.} \bibnamefont{Luttinger}},
  \bibinfo{journal}{Phys. Rev.} \textbf{\bibinfo{volume}{95}},
  \bibinfo{pages}{1154} (\bibinfo{year}{1954}).

\bibitem{jungwirthPRL02}
T.~Jungwirth, Q.~Niu, and A.~H. MacDonald,
\newblock { Phys. Rev. Lett.} \textbf{88},  207208 (2002).



\bibitem[{\citenamefont{Kohn and Luttinger}(1957)}]{kohnPR57}
\bibinfo{author}{\bibfnamefont{W.}~\bibnamefont{Kohn}} \bibnamefont{and}
  \bibinfo{author}{\bibfnamefont{J.~M.} \bibnamefont{Luttinger}},
  \bibinfo{journal}{Phys. Rev.} \textbf{\bibinfo{volume}{108}},
  \bibinfo{pages}{590} (\bibinfo{year}{1957}).

\bibitem[{\citenamefont{Adams and Blount}(1959)}]{adamsJPCS59}
\bibinfo{author}{\bibfnamefont{E.~N.} \bibnamefont{Adams}} \bibnamefont{and}
  \bibinfo{author}{\bibfnamefont{E.~I.} \bibnamefont{Blount}},
  \bibinfo{journal}{J. Phys. Chem. Solids} \textbf{\bibinfo{volume}{10}},
  \bibinfo{pages}{286} (\bibinfo{year}{1959}).

\bibitem[{\citenamefont{Blount}(1962)}]{blountPR62}
\bibinfo{author}{\bibfnamefont{E.~I.} \bibnamefont{Blount}},
  \bibinfo{journal}{Phys. Rev.} \textbf{\bibinfo{volume}{126}},
  \bibinfo{pages}{1636} (\bibinfo{year}{1962}).

\bibitem[{\citenamefont{Chang and Niu}(1995)}]{changPRL95}
\bibinfo{author}{\bibfnamefont{M.-C.} \bibnamefont{Chang}} \bibnamefont{and}
  \bibinfo{author}{\bibfnamefont{Q.}~\bibnamefont{Niu}},
  \bibinfo{journal}{Phys. Rev. Lett.} \textbf{\bibinfo{volume}{75}},
  \bibinfo{pages}{1348} (\bibinfo{year}{1995}).

\bibitem[{\citenamefont{Sundaram and Niu}(1999)}]{sundaramPRB99}
\bibinfo{author}{\bibfnamefont{G.}~\bibnamefont{Sundaram}} \bibnamefont{and}
  \bibinfo{author}{\bibfnamefont{Q.}~\bibnamefont{Niu}},
  \bibinfo{journal}{Phys. Rev. B} \textbf{\bibinfo{volume}{59}},
  \bibinfo{pages}{14915} (\bibinfo{year}{1999}).

\bibitem[{\citenamefont{Haldane}(2004)}]{haldanePRL04}
\bibinfo{author}{\bibfnamefont{F.~D.~M.} \bibnamefont{Haldane}},
  \bibinfo{journal}{Phys. Rev. Lett.} \textbf{\bibinfo{volume}{93}},
  \bibinfo{eid}{206602} (\bibinfo{year}{2004}).

\bibitem[{\citenamefont{Culcer et~al.}(2005)\citenamefont{Culcer, Yao, and
  Niu}}]{culcerPRB05}
\bibinfo{author}{\bibfnamefont{D.}~\bibnamefont{Culcer}},
  \bibinfo{author}{\bibfnamefont{Y.}~\bibnamefont{Yao}}, \bibnamefont{and}
  \bibinfo{author}{\bibfnamefont{Q.}~\bibnamefont{Niu}},
  \bibinfo{journal}{Phys. Rev. B} \textbf{\bibinfo{volume}{72}},
  \bibinfo{pages}{085110} (\bibinfo{year}{2005}).

\bibitem[{\citenamefont{Shindou and Imura}(2005)}]{shindouNPB05}
\bibinfo{author}{\bibfnamefont{R.}~\bibnamefont{Shindou}} \bibnamefont{and}
  \bibinfo{author}{\bibfnamefont{K.-I.} \bibnamefont{Imura}},
  \bibinfo{journal}{Nucl. Phys. B} \textbf{\bibinfo{volume}{720}},
  \bibinfo{pages}{399} (\bibinfo{year}{2005}).
  
\bibitem[{\citenamefont{Xiao et~al.}(2005)\citenamefont{Xiao, Shi, and
  Niu}}]{xiaoPRL05}
\bibinfo{author}{\bibfnamefont{D.}~\bibnamefont{Xiao}},
  \bibinfo{author}{\bibfnamefont{J.}~\bibnamefont{Shi}}, \bibnamefont{and}
  \bibinfo{author}{\bibfnamefont{Q.}~\bibnamefont{Niu}},
  \bibinfo{journal}{Phys. Rev. Lett.} \textbf{\bibinfo{volume}{95}},
  \bibinfo{pages}{137204} (\bibinfo{year}{2005}).

\bibitem[{\citenamefont{Gosselin et~al.}(2006)\citenamefont{Gosselin,
  M{\'e}nas, B{\'e}rard, and Mohrbach}}]{gosselinEPL06}
\bibinfo{author}{\bibfnamefont{P.}~\bibnamefont{Gosselin}},
  \bibinfo{author}{\bibfnamefont{F.}~\bibnamefont{M{\'e}nas}},
  \bibinfo{author}{\bibfnamefont{A.}~\bibnamefont{B{\'e}rard}},
  \bibnamefont{and} \bibinfo{author}{\bibfnamefont{H.}~\bibnamefont{Mohrbach}},
  \bibinfo{journal}{Europhys. Lett.} \textbf{\bibinfo{volume}{76}},
  \bibinfo{pages}{651} (\bibinfo{year}{2006}).

\bibitem[{\citenamefont{Duval et~al.}(2006)\citenamefont{Duval, Horv{\'a}th,
  Horv{\'a}thy, Martina, and Stichel}}]{duvalMPLB06}
\bibinfo{author}{\bibfnamefont{C.}~\bibnamefont{Duval}},
  \bibinfo{author}{\bibfnamefont{Z.}~\bibnamefont{Horv{\'a}th}},
  \bibinfo{author}{\bibfnamefont{P.~A.} \bibnamefont{Horv{\'a}thy}},
  \bibinfo{author}{\bibfnamefont{L.}~\bibnamefont{Martina}}, \bibnamefont{and}
  \bibinfo{author}{\bibfnamefont{P.~C.} \bibnamefont{Stichel}},
  \bibinfo{journal}{Mod. Phys. Lett. B} \textbf{\bibinfo{volume}{20}},
  \bibinfo{pages}{373} (\bibinfo{year}{2006}).

\bibitem[{\citenamefont{Chang and Niu}(2008)}]{changJPCM08}
\bibinfo{author}{\bibfnamefont{M.-C.} \bibnamefont{Chang}} \bibnamefont{and}
  \bibinfo{author}{\bibfnamefont{Q.}~\bibnamefont{Niu}}, \bibinfo{journal}{J.
  Phys.: Condens. Matter} \textbf{\bibinfo{volume}{20}},
  \bibinfo{pages}{193202} (\bibinfo{year}{2008}).

\bibitem[{\citenamefont{Xiao et~al.}(2007)\citenamefont{Xiao, Yao, and
  Niu}}]{xiaoPRL07}
\bibinfo{author}{\bibfnamefont{D.}~\bibnamefont{Xiao}},
  \bibinfo{author}{\bibfnamefont{W.}~\bibnamefont{Yao}}, \bibnamefont{and}
  \bibinfo{author}{\bibfnamefont{Q.}~\bibnamefont{Niu}},
  \bibinfo{journal}{Phys. Rev. Lett.} \textbf{\bibinfo{volume}{99}},
  \bibinfo{eid}{236809} (\bibinfo{year}{2007}).

\bibitem[{\citenamefont{Xiao et~al.}(2006)\citenamefont{Xiao, Yao, Fang, and
  Niu}}]{xiaoPRL06}
\bibinfo{author}{\bibfnamefont{D.}~\bibnamefont{Xiao}},
  \bibinfo{author}{\bibfnamefont{Y.}~\bibnamefont{Yao}},
  \bibinfo{author}{\bibfnamefont{Z.}~\bibnamefont{Fang}}, \bibnamefont{and}
  \bibinfo{author}{\bibfnamefont{Q.}~\bibnamefont{Niu}},
  \bibinfo{journal}{Phys. Rev. Lett.} \textbf{\bibinfo{volume}{97}},
  \bibinfo{eid}{026603} (\bibinfo{year}{2006}).

\bibitem[{\citenamefont{Zhang et~al.}(2009)\citenamefont{Zhang, Tewari, and
  {Das Sarma}}}]{zhangPRB09}
\bibinfo{author}{\bibfnamefont{C.}~\bibnamefont{Zhang}},
  \bibinfo{author}{\bibfnamefont{S.}~\bibnamefont{Tewari}}, \bibnamefont{and}
  \bibinfo{author}{\bibfnamefont{S.}~\bibnamefont{{Das Sarma}}},
  \bibinfo{journal}{Phys. Rev. B} \textbf{\bibinfo{volume}{79}},
  \bibinfo{pages}{245424} (\bibinfo{year}{2009}).

\bibitem[{\citenamefont{Shindou and Balents}(2006)}]{shindouPRL06}
\bibinfo{author}{\bibfnamefont{R.}~\bibnamefont{Shindou}} \bibnamefont{and}
  \bibinfo{author}{\bibfnamefont{L.}~\bibnamefont{Balents}},
  \bibinfo{journal}{Phys. Rev. Lett.} \textbf{\bibinfo{volume}{97}},
  \bibinfo{eid}{216601} (\bibinfo{year}{2006});
  \bibinfo{journal}{Phys. Rev. B} \textbf{\bibinfo{volume}{77}},
  \bibinfo{eid}{035110} (\bibinfo{year}{2008}).
  
    \bibitem{culcerPRB06}
D.~Culcer and Q.~Niu,  { Phys. Rev. B}, \textbf{74}, 035209 (2006).

\bibitem{reichlSP98}
L.~Reichl.
 {\em A Modern Course in Statistical Physics}, (John Wiley and Sons, 1998.);  
J.~Rammer and H.~Smith.
 {\em Rev. Mod. Phys.} \textbf{58}, 323 (1986).




\bibitem[{\citenamefont{Xiao et~al.}(2010)\citenamefont{Xiao, Chang, and
  Niu}}]{xiaoRMP10}
\bibinfo{author}{\bibfnamefont{D.}~\bibnamefont{Xiao}},
  \bibinfo{author}{\bibfnamefont{M.-C.} \bibnamefont{Chang}}, \bibnamefont{and}
  \bibinfo{author}{\bibfnamefont{Q.}~\bibnamefont{Niu}}, \bibinfo{journal}{Rev.
  Mod. Phys.} \textbf{\bibinfo{volume}{82}}, \bibinfo{pages}{1959}
  (\bibinfo{year}{2010}).

\bibitem[{\citenamefont{Murakami and Nagaosa}(2003)}]{murakamiPRL03}
\bibinfo{author}{\bibfnamefont{S.}~\bibnamefont{Murakami}} \bibnamefont{and}
  \bibinfo{author}{\bibfnamefont{N.}~\bibnamefont{Nagaosa}},
  \bibinfo{journal}{Phys. Rev. Lett.} \textbf{\bibinfo{volume}{90}},
  \bibinfo{pages}{057002} (\bibinfo{year}{2003}).

\bibitem[{\citenamefont{Jackiw}(1984)}]{jackiwPRD84}
\bibinfo{author}{\bibfnamefont{R.}~\bibnamefont{Jackiw}},
  \bibinfo{journal}{Phys. Rev. D} \textbf{\bibinfo{volume}{29}},
  \bibinfo{pages}{2375} (\bibinfo{year}{1984}).

\bibitem[{\citenamefont{Hasan and Kane}(2010)}]{hasanRMP10}
\bibinfo{author}{\bibfnamefont{M.~Z.} \bibnamefont{Hasan}} \bibnamefont{and}
  \bibinfo{author}{\bibfnamefont{C.~L.} \bibnamefont{Kane}},
  \bibinfo{journal}{Rev. Mod. Phys.} \textbf{\bibinfo{volume}{82}},
  \bibinfo{pages}{3045} (\bibinfo{year}{2010}).

\bibitem[{\citenamefont{Qi and Zhang}()}]{qiCM10}
\bibinfo{author}{\bibfnamefont{X.-L.} \bibnamefont{Qi}} \bibnamefont{and}
  \bibinfo{author}{\bibfnamefont{S.-C.} \bibnamefont{Zhang}},
  \bibinfo{note}{arXiv:1008.2026}.

\bibitem[{\citenamefont{St{\v{r}}eda}(1982)}]{stredaJPC82}
\bibinfo{author}{\bibfnamefont{P.}~\bibnamefont{St{\v{r}}eda}},
  \bibinfo{journal}{J. Phys. C: Sol. State Phys.}
  \textbf{\bibinfo{volume}{15}}, \bibinfo{pages}{L717} (\bibinfo{year}{1982}).

\bibitem[{\citenamefont{Thouless et~al.}(1982)\citenamefont{Thouless, Kohmoto,
  Nightingale, and {den Nijs}}}]{thoulessPRL82}
\bibinfo{author}{\bibfnamefont{D.~J.} \bibnamefont{Thouless}},
  \bibinfo{author}{\bibfnamefont{M.}~\bibnamefont{Kohmoto}},
  \bibinfo{author}{\bibfnamefont{M.~P.} \bibnamefont{Nightingale}},
  \bibnamefont{and} \bibinfo{author}{\bibfnamefont{M.}~\bibnamefont{{den
  Nijs}}}, \bibinfo{journal}{Phys. Rev. Lett.} \textbf{\bibinfo{volume}{49}},
  \bibinfo{pages}{405} (\bibinfo{year}{1982}).

\bibitem[{\citenamefont{Dirac}(1947)}]{diracBOOK47}
\bibinfo{author}{\bibfnamefont{P.~A.~M.} \bibnamefont{Dirac}},
  \emph{\bibinfo{title}{The Prnciples of Quantum Mechanics}}
  (\bibinfo{publisher}{Oxford University Press}, \bibinfo{year}{1947}).

\bibitem[{\citenamefont{Itzykson and Zuber}(1980)}]{itzyksonBOOK80}
\bibinfo{author}{\bibfnamefont{C.}~\bibnamefont{Itzykson}} \bibnamefont{and}
  \bibinfo{author}{\bibfnamefont{J.-B.} \bibnamefont{Zuber}},
  \emph{\bibinfo{title}{Quantum Field Theory}}
  (\bibinfo{publisher}{McGraw-Hill}, \bibinfo{year}{1980}).

\bibitem[{\citenamefont{Chuu et~al.}(2010)\citenamefont{Chuu, Chang, and
  Niu}}]{chuuSSC10}
\bibinfo{author}{\bibfnamefont{C.-P.} \bibnamefont{Chuu}},
  \bibinfo{author}{\bibfnamefont{M.-C.} \bibnamefont{Chang}}, \bibnamefont{and}
  \bibinfo{author}{\bibfnamefont{Q.}~\bibnamefont{Niu}},
  \bibinfo{journal}{Solid State Commun.} \textbf{\bibinfo{volume}{150}},
  \bibinfo{pages}{533 } (\bibinfo{year}{2010}).

\bibitem[{\citenamefont{Nielsen and Ninomiya}(1983)}]{nielsenPLB83}
\bibinfo{author}{\bibfnamefont{H.~B.} \bibnamefont{Nielsen}} \bibnamefont{and}
  \bibinfo{author}{\bibfnamefont{M.}~\bibnamefont{Ninomiya}},
  \bibinfo{journal}{Phys. Lett. B} \textbf{\bibinfo{volume}{130}},
  \bibinfo{pages}{389} (\bibinfo{year}{1983}).

\bibitem[{\citenamefont{Haldane}(1988)}]{haldanePRL88}
\bibinfo{author}{\bibfnamefont{F.~D.~M.} \bibnamefont{Haldane}},
  \bibinfo{journal}{Phys. Rev. Lett.} \textbf{\bibinfo{volume}{61}},
  \bibinfo{pages}{2015} (\bibinfo{year}{1988}).

\bibitem[{\citenamefont{Foldy and Wouthuysen}(1950)}]{foldyPR50}
\bibinfo{author}{\bibfnamefont{L.~L.} \bibnamefont{Foldy}} \bibnamefont{and}
  \bibinfo{author}{\bibfnamefont{S.~A.} \bibnamefont{Wouthuysen}},
  \bibinfo{journal}{Phys. Rev.} \textbf{\bibinfo{volume}{78}},
  \bibinfo{pages}{29} (\bibinfo{year}{1950}).

\bibitem[{\citenamefont{Rammer and Smith}(1986)}]{rammerRMP86}
\bibinfo{author}{\bibfnamefont{J.}~\bibnamefont{Rammer}} \bibnamefont{and}
  \bibinfo{author}{\bibfnamefont{H.}~\bibnamefont{Smith}},
  \bibinfo{journal}{Rev. Mod. Phys.} \textbf{\bibinfo{volume}{58}},
  \bibinfo{pages}{323} (\bibinfo{year}{1986}).

\bibitem[{\citenamefont{Wigner}(1932)}]{wignerPR32}
\bibinfo{author}{\bibfnamefont{E.}~\bibnamefont{Wigner}},
  \bibinfo{journal}{Phys. Rev.} \textbf{\bibinfo{volume}{40}},
  \bibinfo{pages}{749} (\bibinfo{year}{1932}).

\bibitem[{\citenamefont{Weinberg}(1996)}]{weinbergBook2}
\bibinfo{author}{\bibfnamefont{S.}~\bibnamefont{Weinberg}},
  \emph{\bibinfo{title}{The Quantum Theory of Fields}},
  vol.~\bibinfo{volume}{2} (\bibinfo{publisher}{Cambridge University Press},
  \bibinfo{year}{1996}).
  
\bibitem[{\citenamefont{Eguchi et~al.}(1980)\citenamefont{Eguchi, Gilkey, and
  Hanson}}]{eguchiPRP80}
\bibinfo{author}{\bibfnamefont{T.}~\bibnamefont{Eguchi}},
  \bibinfo{author}{\bibfnamefont{P.~B.} \bibnamefont{Gilkey}},
  \bibnamefont{and} \bibinfo{author}{\bibfnamefont{A.~J.}
  \bibnamefont{Hanson}}, \bibinfo{journal}{Phys. Rep.}
  \textbf{\bibinfo{volume}{66}}, \bibinfo{pages}{213} (\bibinfo{year}{1980}).

\bibitem[{\citenamefont{Wong and Tserkovnyak}(2009)}]{wongPRB09}
\bibinfo{author}{\bibfnamefont{C.~H.} \bibnamefont{Wong}} \bibnamefont{and}
  \bibinfo{author}{\bibfnamefont{Y.}~\bibnamefont{Tserkovnyak}},
  \bibinfo{journal}{Phys. Rev. B} \textbf{\bibinfo{volume}{80}},
  \bibinfo{eid}{184411} (\bibinfo{year}{2009}).

\bibitem[{\citenamefont{Semenoff}(1984)}]{semenoffPRL84}
\bibinfo{author}{\bibfnamefont{G.~W.} \bibnamefont{Semenoff}},
  \bibinfo{journal}{Phys. Rev. Lett.} \textbf{\bibinfo{volume}{53}},
  \bibinfo{pages}{2449} (\bibinfo{year}{1984}).

\bibitem[{\citenamefont{Redlich}(1984)}]{redlichPRL84}
\bibinfo{author}{\bibfnamefont{A.}~\bibnamefont{Redlich}},
  \bibinfo{journal}{Phys. Rev. Lett.}
  \textbf{\bibinfo{volume}{52}}, \bibinfo{pages}{18} (\bibinfo{year}{1984}).

\bibitem[{\citenamefont{Bransden and Joachain}(2000)}]{bransdenBOOK00}
\bibinfo{author}{\bibfnamefont{B.~H.} \bibnamefont{Bransden}} \bibnamefont{and}
  \bibinfo{author}{\bibfnamefont{C.~J.} \bibnamefont{Joachain}},
  \emph{\bibinfo{title}{Quantum Mechanics}} (\bibinfo{publisher}{Prentice
  Hall}, \bibinfo{year}{2000}), \bibinfo{edition}{2nd} ed.

\end{thebibliography}
\end{document}